\documentclass[sigconf]{acmart}
\AtBeginDocument{%
  }

\setcopyright{acmlicensed}
\copyrightyear{2026}
\acmYear{2026}
\acmDOI{XXXXXXX.XXXXXXX}
\acmConference[Conference acronym 'XX]{Make sure to enter the correct
  conference title from your rights confirmation email}{June 03--05,
  2018}{Woodstock, NY}

\usepackage{subcaption}
\usepackage{amsmath,amssymb,amsfonts}
\usepackage{float}
\usepackage{rotating}

\usepackage{bm}\usepackage{bm}
\usepackage{booktabs} 
\usepackage{multirow,xspace,pifont}
\usepackage{graphicx}
\usepackage{textcomp}
\usepackage{tcolorbox}
\usepackage{siunitx}
\usepackage{mdframed}
\usepackage{paralist}
\usepackage{listings,verbatim,cleveref}
\usepackage{caption}
\usepackage{makecell}
\usepackage{xspace}
\usepackage{fontawesome5} 
\usepackage{algorithm}
\usepackage{algpseudocode}

\usepackage{tikz}
\usetikzlibrary{arrows.meta,positioning,calc,shapes.geometric}

\usepackage[utf8]{inputenc} 
\usepackage[T1]{fontenc}    
\usepackage{url}            
\usepackage{nicefrac}       
\usepackage{microtype}      
\usepackage{xcolor}         

\usepackage{todonotes}
\newcommand{\lb}{\ensuremath{\bm{l}}}
\newcommand{\ub}{\ensuremath{\bm{u}}}

\newcommand{\seedpool}{\ensuremath{\mathcal{P}}}

\newcommand{\iso}{\textsc{Batch\text{-}Iso}\xspace}
\newcommand{\ani}{\textsc{Batch\text{-}Ani}\xspace}
\newcommand{\fix}{\textsc{Seq\text{-}Fixed}\xspace}

\definecolor{myred}{RGB}{255, 204, 204}   
\definecolor{mygreen}{RGB}{204, 255, 204} 
\newcommand{\myparagraph}[1]{\par\addvspace{\baselineskip}\noindent{\bf #1.}}
\newcommand{\revision}[1]{\textcolor{black}{#1}}
\setcopyright{none}
\renewcommand\footnotetextcopyrightpermission[1]{}

\begin{document}

\title{Tensor-Based Batch Fuzzing with Adaptive Perturbation Scaling for Deep Neural Networks}

\author{Guanqin Zhang}
\orcid{0000-0002-3844-8180}
\affiliation{%
  \institution{University of New South Wales}
  \city{Sydney}
  \country{Australia}}
\email{guanqin.zhang@unsw.edu.au}

\author{Yulei Sui}
\orcid{0000-0002-9510-6574}
\affiliation{%
  \institution{University of New South Wales}
  \city{Sydney}
  \country{Australia}}
\email{y.sui@unsw.edu.au}

\renewcommand{\shortauthors}{Guanqin Zhang and Yulei Sui}

\begin{abstract}
Deep neural networks (DNNs) are increasingly deployed in safety-critical domains such as autonomous driving and medical diagnosis, yet their opaque, high-dimensional parameter spaces make it difficult to systematically assess model reliability on unseen inputs. Existing coverage-guided sequential fuzzing frameworks for DNN inherit a one-input-per-iteration design from traditional software fuzzing and apply uniform perturbation budgets across all input dimensions, limiting both testing throughput (i.e., inputs processed per unit time) and the precision of input-space exploration. 

We present a new specification-aware batch fuzzing framework with adaptive perturbation scaling that addresses both limitations. Rather than relying on a fixed global perturbation radius $\epsilon$, our approach derives mutation step sizes from specification-defined feasible ranges (i.e., the gap between lower and upper bounds) using a shared scale factor. This scaling can be applied either as a global scalar (\emph{isotropic}) or as per-dimension step sizes (\emph{anisotropic}), enabling perturbations to remain consistent with the underlying constraint structure. As a result, the fuzzer can explore input spaces with heterogeneous feature scales more effectively across all specifications in the batch.
We embed input constraints and output property checks directly into the network as non-trainable layers, yielding a wrapped model that processes $B$ specification instances in a single batched iteration, which substantially improves fuzzing efficiency and exploration of counterexamples.

We evaluate our framework extensively on three benchmarks, covering six networks and over 400 specifications across \textsc{TrafficSigns}, \textsc{Cifar100}, and \textsc{TinyImageNet}. Our tensor-based fuzzing achieves up to 40$\times$ higher throughput and 4$\times$ more violations than the sequential baseline under the same time budget, demonstrating significantly improved effectiveness in specification-guided fuzzing.
\end{abstract}




\maketitle

\section{Introduction}
DNNs have achieved strong performance across image classification, object detection, and natural language understanding, and are increasingly deployed in safety-critical domains such as autonomous driving~\cite{bojarski2016end}, medical diagnosis, and cyber-physical control~\cite{amodei2016concrete,dalrymple2024towards}, where incorrect predictions can have severe robustness issues. This trend has prompted growing attention toward establishing rigorous assurance for DNN-based systems~\cite{mitra2024formal,boudardara2024review}. 

\myparagraph{\text{Existing efforts}} Coverage-guided grey- or white-box fuzzing has emerged as the new paradigm for assessing DNN reliability, adapting techniques from traditional software fuzzing~\cite{manes2019art,bohme2016coverage}. The fuzzing process revolves around three key elements: a \emph{coverage metric} that quantifies how thoroughly a test suite exercises the DNN internal computation, a \emph{test case selection} strategy that prioritizes promising inputs (high energy inputs), and a \emph{mutation strategy} that generates new candidate inputs. 
These elements form an iterative loop in which selected seeds are mutated, fed through the network, and evaluated against the coverage metric; candidates that increase coverage are retained for subsequent rounds. 
DeepXplore~\cite{pei2017deepxplore} introduced \emph{neuron coverage} (NC) as the white-box adequacy criterion, measuring the fraction of neurons whose activation exceeds a threshold, and DeepGauge~\cite{ma2018deepgauge} extended this to multi-granularity criteria. Building on these metrics, DeepHunter~\cite{xie2019deephunter} applies metamorphic mutation under coverage feedback, TensorFuzz~\cite{odena2019tensorfuzz} uses approximate nearest-neighbour coverage, DeepTest~\cite{tian2018deeptest} generates domain-specific transformations, and DLFuzz~\cite{guo2018dlfuzz} maximizes coverage via gradient-guided differential fuzzing.

\myparagraph{\text{Limitations}} 
Existing DNN fuzzing methods like DeepHunter~\cite{xie2019deephunter} suffer from two key limitations. 
First, they are fundamentally organized around a \textit{one-input-per-iteration} fuzzing paradigm from traditional software fuzzing, where each mutated input is executed independently~\cite{bohme2016coverage,miller1990empirical}. 
However, DNN inference is inherently tensorized: a forward pass natively supports batched evaluation, allowing $B$ inputs to be processed with nearly the same kernel-launch overhead as a single input, with computation amortized by hardware-level parallelism~\cite{paszke2019pytorch,abadi2016tensorflow}. 
While TensorFuzz~\cite{odena2019tensorfuzz}, which is designed for TensorFlow programs, exploits batching for model evaluation, its core fuzzing operators (NumPy-based
mutation and per-element coverage feedback) are not tensor-based transformations (e.g., PyTorch tensors) across the batch, resulting in underutilized throughput.
Second, most approaches adopt a fixed mutation strategy (i.e., a uniform perturbation $\epsilon$) across all input dimensions~\cite{goodfellow2015explaining,madry2017towards}. 
In practice, each dimension $d \in D$ may correspond to a pixel intensity or a physical feature, and can be constrained by its own interval $[l_d, u_d]$ defined in the input specification. 
Formal formats such as VNNLib~\cite{brix2023first,2024vnncomp} explicitly support such non-uniform, per-dimension bounds. 
Ignoring this structure forces the fuzzer to either under-explore loosely constrained dimensions or over-explore tightly constrained ones, leading to inefficient search.

\begin{table*}[t]
\centering
\caption{Comparison of Fuzzing approaches.
Lower and upper bounds: $\lb, \ub \in \mathbb{R}^{B \times D}$ with batch size $B$, input dim $D$ (e.g., $D\!=\!784$ for MNIST), output dim $D'$.
Mutation: $\eta \cdot \Delta$ (uniform scalar) or $\mathcal{S} \odot \Delta$
(per-element; $\odot$: Hadamard product), where $\Delta \in \mathbb{R}^{B \times D}$ is the
mutation direction matrix whose $b$-th row equals the per-sample direction
$\bm{\delta}^{(b)}$ (FGSM, PGD, etc.), and $\bm{\delta}\!=\!\Delta$ when $B$=1.
Mean range: $\bar{r}\! =\! \frac{1}{BD} \sum_{b,d} (u_{b,d} - l_{b,d})$ across all samples and dimensions.
Uniform scaling: $\eta\!=\! \bar{r} \cdot s$.
Per-element scaling tensor: $\mathcal{S}_{b,d} = (u_{b,d} - l_{b,d}) \cdot s$ gives each
sample-dimension pair its own perturbation magnitude, where parameter $s\! \in\! (0,1]$ adaptively controls
granularity (e.g., $s=0.1$ means each step moves $10\%$ of the feasible range,
requiring $\sim$10 steps to traverse from $\lb$ to $\ub$).
}
\label{tab:fuzzing-comparison}
\begin{tabular}{l@{\hspace{8mm}}c@{\hspace{8mm}}c@{\hspace{8mm}}c}
\toprule
\textbf{Aspect} & \textbf{Sequential} & \textbf{Batch Isotropic (Ours)} & \textbf{Batch Anisotropic (Ours)} \\
\midrule
Seed selection & $\bm{x} \in \mathbb{R}^D$ & $\bm{X} \in \mathbb{R}^{B \times D}$ & $\bm{X} \in \mathbb{R}^{B \times D}$ \\ 
Perturbation value& $\eta \in \mathbb{R}$ & $\eta \in \mathbb{R}$ & $\mathcal{S} \in \mathbb{R}^{B \times D}$ \\
Perturbation scaling& Fixed \ (e.g., $\eta\!=\!0.01$) & Adaptive scalar \ ($\eta = \bar{r} \cdot s$) 
& Adaptive tensor \ ($\mathcal{S}_{b,d} = (\ub_{b,d} - \lb_{b,d}) \cdot s$)\\
Mutation & $\tilde{\bm{x}} = \bm{x} + \eta \cdot \bm{\delta}$ & $\tilde{\bm{X}} = \bm{X} + \eta \cdot \Delta$ & $\tilde{\bm{X}} = \bm{X} + \mathcal{S} \odot \Delta$ \\
Inference & $f: \mathbb{R}^D \to \mathbb{R}^{D'}$ & $f: \mathbb{R}^{B \times D} \to \mathbb{R}^{B \times D'}$ & $f: \mathbb{R}^{B \times D} \to \mathbb{R}^{B \times D'}$ \\ \midrule
Samples per iteration & 1  & $B$  & $B$  \\
Bounds handling & Single $[\lb, \ub]$ / fixed & Shared $[\lb, \ub]$ / uniform & Per-sample $[\lb_b, \ub_b]$ / non-uniform \\
Spec heterogeneity & Homogeneous & Homogeneous & Heterogeneous \\
\bottomrule
\label{tab:diff}
\end{tabular}
\end{table*}

\myparagraph{\text{Our approach}} 
To address the above limitations, we introduce specification-aware batch fuzzing, shifting DNN fuzzing from sequential input-level mutation to batch specification-level execution. Fuzzing is lifted to the specification level, not the input level, with the entire loop expressed in tensor-level semantics to avoid sequential per-input processing in prior works~\cite{xie2019deephunter,odena2019tensorfuzz}.
We embed specification constraints directly into the network as non-trainable layers, yielding a wrapped model $M_{\mathrm{wrapped}}$ that integrates constraints and property checking into standard forward execution. The model encodes per-sample input bounds $[l_b, u_b] \in \mathbb{R}^D$ alongside output specifications, enabling batched inputs $X \in \mathbb{R}^{B \times D}$ to be processed in a single pass with per-sample verdicts. These bounds are reused throughout the pipeline for projection, adaptive perturbation scaling and validity checking, providing a specification-aware batch pipeline to significantly improve fuzzing effectiveness.

Table~\ref{tab:diff} compares sequential and batch fuzzing.
Our formulation lifts all phases of the fuzzing loop (i.e., seed selection, mutation, inference, coverage tracking, and feedback) to operate over a $B$-sample batch within a single iteration, where each sample (batch element) corresponds to a specification instance consisting of an input and its associated input constraints and output property. This effectively leverages the batch dimension as a throughput multiplier without introducing additional algorithmic approximations.

To address the fixed perturbation limitation, we introduce \emph{adaptive perturbation scaling} which derives mutation magnitudes from specification-defined feasible ranges via a shared scale factor $s$. 
Our approach supports both \emph{isotropic scaling}, where a global step size is adaptively derived from feasible ranges, and \emph{anisotropic scaling}, which assigns per-dimension step sizes from these ranges, enabling consistent exploration under heterogeneous input constraints.

As shown in Table~\ref{tab:fuzzing-comparison}, the three paradigms differ primarily in bounds handling and perturbation scaling. 
Sequential fuzzing applies a fixed step size $\eta$ to individual samples without adaptation. 
Batch isotropic fuzzing shares bounds $[\lb, \ub]$ across $B$ samples and computes a scalar step size $\eta = \bar{r} \cdot s$ from the mean range $\bar{r}$, broadcasting uniform perturbations across all dimensions. 
In contrast, batch anisotropic fuzzing maintains per-element bounds $[\lb, \ub] \in \mathbb{R}^{B \times D}$ and computes a scaling tensor $\mathcal{S}_{b,d} = (\ub_{b,d} - \lb_{b,d}) \cdot s$, assigning larger steps to loosely constrained dimensions and finer steps to tightly constrained ones. 
The ``spec heterogeneity'' row highlights that the anisotropic formulation supports different specifications per sample within the same batch.

\noindent\textbf{Our contributions} are summarized as below:

\begin{compactitem}
\item We introduce a specification-aware batch fuzzing framework that operates at the level of specification instances rather than individual inputs. The full fuzzing loop, including seed selection, mutation, inference, coverage tracking, and feedback, is expressed as a unified data-parallel pipeline. By embedding input bounds and output checks as non-trainable layers in a wrapped model, specifications are evaluated in a single forward pass, significantly improving fuzzing efficiency and specification-guided exploration.

\item We propose an \emph{adaptive perturbation scaling} scheme that derives mutation magnitudes from specification-defined bounds. It supports two scaling methods: isotropic scaling, where a global step size is adaptively derived from feasible ranges, and anisotropic scaling, which assigns per-dimension step sizes from the same ranges, enabling consistent exploration under heterogeneous input constraints.

\item We implement the framework in PyTorch and evaluate it extensively on six networks with over 400 specifications, demonstrating significant performance gains and achieving up to 40$\times$ higher throughput than the sequential baseline. Our tensor-based batch fuzzing processes all $B$ specifications simultaneously within a single 60\,s window and discovers 4$\times$ more violations, whereas the sequential baseline requires $B \times 60$,s to cover the same specifications.
\end{compactitem}

\section{Preliminaries and Sequential DNN Fuzzing}
\label{sec:prelim}

\myparagraph{Neural Network}
A feed-forward neural network $f\colon\mathbb{R}^n\to\mathbb{R}^{n'}$
transforms an $n$-dimensional input through $k\in K$ successive layers to
produce an $n'$-dimensional output.
For our fuzzing, we treat $f$ as a \emph{grey-box} network under test: gradient-driven mutation
operators (PGD, FGSM) require differentiable access via the backward
pass, while heuristic operators (\textsc{Random}, \textsc{Boundary})
require only forward evaluation. In either case, no knowledge of the training procedure or dataset beyond the model weights is assumed.

\myparagraph{Specification and Property Violation}
A \emph{specification} $(\Phi, \Psi)$ pairs an \emph{input
constraint} $\Phi$ with an \emph{output property}
$\Psi\colon\mathbb{R}^{n'}\to\{\textit{true},\textit{false}\}$.
The feasible perturbation region induced by $\Phi$ takes the form of
a box constraint
\begin{align}\label{eq:inputspec}
  \llbracket\Phi\rrbracket
    := \bigl\{\bm{x}\in\mathbb{R}^n
       \mid \lb\leq\bm{x}\leq\ub\bigr\},
\end{align}
where $\lb$ and $\ub$ are
given directly in explicit box specification (e.g., per-dimension VNNLib
bounds~\cite{brix2023first,2024vnncomp}). $\lb=\bm{oi}-\varepsilon$, $\ub=\bm{oi}+\varepsilon$ can also be derived from  $\ell_\infty$-balls specifications with an original image $\bm{oi}$ and perturbation radius $\varepsilon$.
The output property $\Psi$ requires that the predicted
class remains unchanged for all inputs within
$\llbracket\Phi\rrbracket$:
\begin{align}\label{eq:outputspec}
  \Psi(f(\bm{x})) = \textit{true}
  \iff
  \min_{c \neq c^*}
  \bigl(f(\bm{x})_{c^*} - f(\bm{x})_c\bigr) > 0,
\end{align}
where $c^* = \arg\max_c f(\bm{x}_0)_c$ is the reference predicted
class with the maximum score, and $c$ is any other candidate class and $f(\bm{x}_0)_c$ is the score/logit for class $c$. More generally, $\Psi$ can encode any linear inequality over
the output, accommodating margin and range
constraints~\cite{geng2023towards,surveyliu}.
One of the main objectives of fuzzing is to find a \emph{counterexample}: a
candidate input $\tilde{\bm{x}}\in\llbracket\Phi\rrbracket$ such
that $\Psi(f(\tilde{\bm{x}}))=\textit{false}$, constituting a
concrete \emph{property violation} (e.g., an incorrectly predicted class).

\myparagraph{Sequential DNN Fuzzing}
Algorithm~\ref{alg:sequential} presents a classical one-input-per-iteration
fuzzing loop that underpins existing tools such as
DeepHunter~\cite{xie2019deephunter} and
TensorFuzz~\cite{odena2019tensorfuzz}.

\begin{algorithm}[htbp]
\caption{Sequential DNN Fuzzing (baseline)}\label{alg:sequential}
\small
\begin{algorithmic}[1]
\Require Network $f$; specification $(\Phi, \Psi)$; timeout $t_{\max}$
\Ensure  Counterexample set $\mathcal{C}$;
         coverage $\mathrm{Cov}_{\mathrm{global}}$
\Statex \textcolor{green!60!black}{$/*$\textit{Initialization}$*/$}
\State $\seedpool \gets \{(\bm{x}_0, e_{0})\};\;
       \mathcal{C} \gets \emptyset;\; t \gets 0$
\State $\forall k:\! \bm{m}_k \!\gets\! \bm{0}^{d_k};\;
       \bm{m} \!\gets\! \mathrm{concat}_k(\bm{m}_k);$
\hfill $\triangleright$ coverage $\bm{m}_k$ for layer $k$
\While{$\textsc{Clock}() < t_{\max}$}
  \Statex \textcolor{green!60!black}{$/*$\textit{Phase 1 -- Seed Selection}$*/$}
  \State $\bm{x} \gets \textsc{SeedSelect}(\seedpool)$
         \hfill$\triangleright$ one seed $\in \mathbb{R}^{n}$
  \Statex \textcolor{green!60!black}{$/*$\textit{Phase 2 -- Mutation (under input constraints)}$*/$}
  \State $\tilde{\bm{x}} \gets
    \Pi_{\llbracket\Phi\rrbracket}\bigl(\bm{x} + \eta \cdot \bm{\delta}\bigr)$
    \hfill $\triangleright$ perturbation direction $\bm{\delta}$; step size $\eta$
  \Statex \textcolor{green!60!black}{$/*$\textit{Phase 3 -- Execution}$*/$}
  \State $A, \tilde{\bm{y}} \gets f(\tilde{\bm{x}})$
         \hfill$\triangleright$ activation map $A$ with one mutant
  \Statex \textcolor{green!60!black}{$/*$\textit{Phase 4 -- Feedback and update}$*/$}
  \State $q, \bm{m} \gets \textsc{CoverageUpdate}(A,\bm{m})$
         \hfill$\triangleright$ coverage gain $q\!\in\!\{0,1\}$
  \State $v \gets \bigl[\Psi(\tilde{\bm{y}}) = \textit{false}\bigr]$
         \hfill$\triangleright$ spec violation $v\!\in\!\{0,1\}$
\State $e \gets q\alpha + v\beta$
\hfill $\triangleright$ coverage/violation reward weights $\alpha$/$\beta$
\State \textbf{if} $q \lor v$ \textbf{then}\; $\seedpool.\textsc{Insert}(\tilde{\bm{x}},\, \textit{energy}=e)$
  \State \textbf{if} $v$ \textbf{then}\; $\mathcal{C} \gets \mathcal{C} \cup \{\tilde{\bm{x}}\}$
\EndWhile
\State $\mathrm{Cov}_{\mathrm{global}}
      \gets \frac{\sum_k \|\bm{m}_k\|_0}{\sum_k d_k}$ \hfill $\triangleright$ $\bm{m}_k \!\in\! \{0,1\}^{d_k}$; $d_k$:  neuron num at layer $k$
\State \Return $\mathcal{C},\; \mathrm{Cov}_{\mathrm{global}}$
\end{algorithmic}
\end{algorithm}

The loop revolves around a mutable \emph{seed corpus} $\seedpool$, initialized with $(\bm{x}_0, e_0)$ (Line 1), where $\bm{x}_0$ is the original seed input and $e_0 \in \mathbb{R}{>0}$ is its initial energy. Each seed $\bm{x}$ defines a feasible perturbation region $\llbracket \Phi \rrbracket$ within which mutations are confined. Seeds are sampled proportionally to their energy based on feedback from the coverage bitmap $\bm{m}$ (Line 2), and new candidates are added to $\seedpool$ if they increase neuron coverage or trigger a property violation (Line 10).

At each iteration, a single seed $\bm{x}$ is drawn from the corpus
proportionally to its energy (Line~4), a perturbation $\bm{\delta}$
is applied and projected onto the feasible region
$\llbracket\Phi\rrbracket$ (Line~5), and the mutated input is
forwarded through $f$ to obtain the output and intermediate
activations $A$ (Line~6). 
The perturbation direction $\bm{\delta}$ can be
constructed via gradient-driven methods such as Projected Gradient Descent (PGD)~\cite{madry2017towards}:
\begin{align}\label{eq:pgd}
  \bm{\delta} = \alpha \cdot \mathrm{sign} \bigl(
  \nabla_{\bm{x}}\, \mathcal{L}(f(\bm{x}), y)\bigr)
\end{align}
where $\mathcal{L}$ is the classification loss, $y$ is the
ground-truth label, and $\alpha \leq \eta$ is the step size;
alternatively, $\bm{\delta}$ may be sampled from a heuristic
distribution such as uniform~\cite{tian2018deeptest} or Gaussian
noise~\cite{odena2019tensorfuzz}.

The quality of mutated inputs is evaluated via \emph{neuron
coverage}~\cite{pei2017deepxplore,ma2018deepgauge} (Line~7).
During the forward pass, PyTorch hooks~\cite{paszke2019pytorch} intercept intermediate activations $A_k$ at each layer $k$ without modifying the computation graph; spatial activations (convolutional layers) are reduced
per-channel via absolute-value max-pooling, while fully connected layers use $A_k[j]$ directly, where $j$ is the neuron index within layer $k$.
Two coverage strategies are typically supported.
\emph{Global coverage} maintains a per-layer bitmap
$\bm{m}_k \in \{0,1\}^{d_k}$ recording which neurons have ever fired (i.e., $|A_k[j]|>\tau$, threshold $\tau>0$)~\cite{pei2017deepxplore,ma2018deepgauge}:
\begin{align}\label{eq:coverage}
  \mathrm{Cov}_{\mathrm{global}}
    = \frac{\sum_k \|\bm{m}_k\|_0}{\sum_k d_k}.
\end{align}
\emph{Best-input coverage} requires no persistent
bitmap; it tracks the running maximum of the per-input fraction
$\rho(\tilde{\bm{x}}) = \tfrac{1}{D}|\{(k,j)\mid|A_k[j]|>\tau\}|$,
reporting $\mathrm{Cov}_{\mathrm{global}} = \max_{\tilde{\bm{x}}} \rho(\tilde{\bm{x}})$.
In both cases, $q=1$ if $\tilde{\bm{x}}$ strictly improves the current
coverage state.
Coverage serves as
the structural feedback signal that guides corpus growth: an input
that activates a previously uncovered neuron is treated as
interesting and retained for subsequent mutation, analogous to the
new-edge criterion in traditional coverage-guided
fuzzers~\cite{manes2019art}. Line~8 checks whether the output
property $\Psi$ is violated. Seeds satisfying either criterion are
inserted into the corpus with updated energy (lines~9--10), and any
violating inputs are recorded as counterexamples (lines~11).

\section{Overview of Our Batch DNN Fuzzing}
\label{sec:overview}

\begin{figure*}[!ht]
    \centering
    \includegraphics[width=\linewidth]{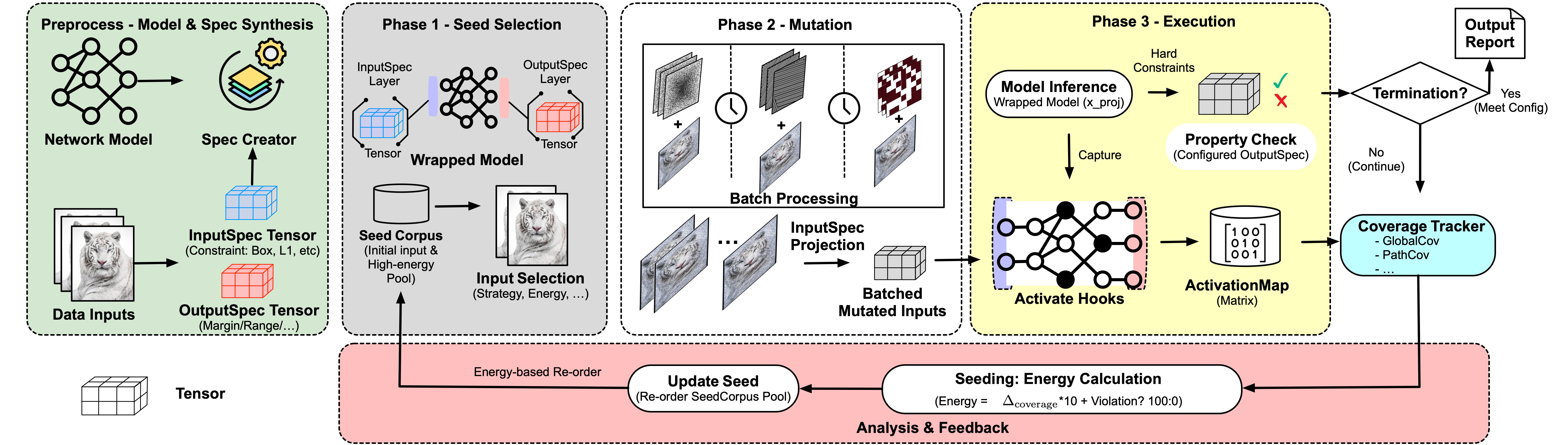}
    \caption{Overview of our tensor-based batch fuzzing framework, which processes specification instances in parallel across four phases (i.e., seed selection, mutation, execution, and  feedback), each operating on batched tensors of size $B$.}
    \label{fig:overview}
\end{figure*}

We first present a high-level overview of our batch fuzzing approach, as depicted in Figure~\ref{fig:overview}. We then provide the overview algorithm (Algorithm~\ref{alg:overall}) for batch fuzzing with isotropic and anisotropic perturbations, which contrasts directly with the sequential version (Algorithm~\ref{alg:sequential}). Each phase is then detailed in Section~\ref{sec:methods}, including its internal workflows and associated sub-algorithms.

Figure~\ref{fig:overview} illustrates the four-phase fuzzing pipeline.
The \emph{preprocessing} stage synthesises a wrapped model by composing
the network with an \texttt{InputSpecLayer} and \texttt{OutputSpecLayer},
encoding the input constraint $\Phi$ and output property $\Psi$ as
tensor-native layers alongside the initial seed corpus.
In \emph{Phase~1}, seeds are drawn from the corpus by energy-weighted
sampling, prioritising inputs that previously triggered new coverage or
property violations.
In \emph{Phase~2}, the selected batch is perturbed and projected onto
the feasible region defined by $\Phi$, producing batched mutated inputs
$\tilde{X}$ that satisfy the hard input constraints by construction.
\emph{Phase~3} runs a single batched forward pass: PyTorch activation
hooks capture the layer-wise activation map $A$, while the
\texttt{OutputSpecLayer} performs a vectorised property check, flagging
samples that violate $\Psi$.
The \emph{analysis and feedback} loop then computes per-sample energy
$e = \Delta_{\mathrm{cov}} \cdot \alpha + \mathbf{1}[\text{violation}] \cdot \beta$,
updates the coverage tracker, and re-orders the seed corpus accordingly; the loop repeats until the termination condition is met, after which an output report summarizing counterexamples and final coverage is produced.

Algorithm~\ref{alg:overall} presents the overview algorithm for batch fuzzing, formalising the four phases shown in Figure~\ref{fig:overview}. Its tensor-based data structures and adaptive perturbation scaling contrast with the sequential algorithm introduced earlier in Algorithm~\ref{alg:sequential}.
The key differences are threefold.
First, seed selection (Line~4) draws a batch $X \in \mathbb{R}^{B \times D}$
in a single call rather than one seed at a time, with all $B$ samples
mutated in parallel.
Second, Lines~5--6 introduce a dedicated perturbation scaling step that replaces the fixed scalar $\eta$: under \textsf{isotropic} mode a single scalar $\eta = \bar{r} \cdot s$ is adaptively derived from the mean feasible range $\bar{r}$, while under \textsf{anisotropic} mode a per-element tensor $\mathcal{S} \in \mathbb{R}^{B \times D}$ with $\mathcal{S}_{b,d} = (\ub_{b,d} - \lb_{b,d}) \cdot s$ assigns each
sample--dimension pair its own perturbation magnitude; both modes apply the same mutation step at line~7, $\tilde{X} = \Pi_{\llbracket\Phi\rrbracket}(X + \eta \odot \Delta)$,
where $\odot$ reduces to scalar multiplication in the isotropic case.
Third, lines~9--13 lift the feedback variables $\bm{q}$, $\bm{v}$,
and $\bm{e}$ from scalars to length-$B$ vectors, with energy assignment and corpus insertion applied element-wise across the batch. All other algorithmic structures, including coverage bitmap update (line~9), spec violation check (line~10), and the final
$\mathrm{Cov}_{\mathrm{global}}$ formula (line~15), are identical to
the sequential case.

\begin{algorithm}[t]
\caption{Batch DNN Fuzzing (Isotropic \& Anisotropic)}\label{alg:overall}
\small
\begin{algorithmic}[1]
\Require Network $f$; specification $(\Phi, \Psi)$; timeout $t_{\max}$;
         batch size $B$; strategy weights $W$; reward weights $\alpha,\beta$;
         activation threshold $\tau$; minimum energy $e_{\min}$;
         perturbation mode $\pi \in \{\mathrm{\textsf{isotropic}},\mathrm{\textsf{anisotropic}}\}$;
         granularity $s \in (0,1]$
\Ensure  Counterexample set $\mathcal{C}$;
         coverage $\mathrm{Cov}_{\mathrm{global}}$
\Statex \textcolor{green!60!black}{$/*$\textit{Initialization}$*/$}
\State $\seedpool \gets \{(\bm{x}_0,\, e_0)\};\;
       \mathcal{C} \gets \emptyset$
       \hfill$\triangleright$ $e_0 = 1$
\State $\forall k:\bm{m}_k \!\gets\! \bm{0}^{d_k};\;
       \bm{m} \!\gets\! \mathrm{concat}_k(\bm{m}_k)$
       \hfill$\triangleright$ $\bm{m}_k \!\in\! \{0,1\}^{d_k}$: coverage map
\While{$\textsc{Clock}() < t_{\max}$}
  \Statex \textcolor{green!60!black}{$/*$\textit{Phase 1 -- Seed Selection}$*/$}
  \State $\bm{X} \gets \textsc{SeedSelect}(\seedpool,\, B)$
         \hfill$\triangleright$ $\bm{X} \in \mathbb{R}^{B \times D}$, sampled $\propto$ energy
  \Statex \textcolor{green!60!black}{$/*$\textit{Phase 2 -- Mutation}$*/$}
  \State $s \gets 1 - (1 - s)^{\,n + 1}$
         \hfill$\triangleright$ $n$: cumulative selection count of seed
  \State $\eta \gets
    \begin{cases}
      \bar{r} \cdot s \in \mathbb{R},\ \text{where}\ \bar{r} = \tfrac{1}{BD}\textstyle\sum_{b,d}(\ub_{b,d}-\lb_{b,d})
        & \pi = \mathrm{\textsf{isotropic}} \\[4pt]
      \mathcal{S} \in \mathbb{R}^{B \times D},\;
      \mathcal{S}_{b,d} = (\ub_{b,d}-\lb_{b,d})\cdot s
        & \pi = \mathrm{\textsf{anisotropic}}
    \end{cases}$
\State $\tilde{\bm{X}} \gets \Pi_{\llbracket\Phi\rrbracket}(\bm{X} + \eta \odot \Delta)$
       \hfill$\triangleright$ $\Delta \in \mathbb{R}^{B \times D}$: direction matrix;\; $\odot$ broadcasts when $\eta \in \mathbb{R}$
  \Statex \textcolor{green!60!black}{$/*$\textit{Phase 3 -- Execution}$*/$}
  \State $(\hat{\bm{Y}},\, A) \gets f(\tilde{\bm{X}})$
         \hfill$\triangleright$ batched forward inference;\; $\hat{\bm{Y}} \in \mathbb{R}^{B \times D'}$
  \Statex \textcolor{green!60!black}{$/*$\textit{Phase 4 -- Feedback and update}$*/$}
  \State $\bm{q},\, \bm{m} \gets \textsc{CoverageUpdate}(A,\, \bm{m})$
         \hfill$\triangleright$ $\bm{q} \in \{0,1\}^B$: per-sample coverage gain
  \State $\bm{v} \gets \bigl[\Psi(\hat{\bm{Y}}) = \mathit{false}\bigr]$
         \hfill$\triangleright$ $\bm{v} \in \{0,1\}^B$: per-sample spec violation
  \State $\bm{e} \gets \max(\bm{q}\alpha + \bm{v}\beta,\; e_{\min})$
         \hfill$\triangleright$ element-wise;\; $\bm{e} \in \mathbb{R}^B$
  \State \textbf{for} $b$\ \textbf{where}\ $q_b \lor v_b$:\;
         $\seedpool.\textsc{Insert}(\tilde{\bm{X}}[b],\; \mathit{energy}=e_b)$
  \State $\mathcal{C} \gets \mathcal{C} \cup \{\tilde{\bm{X}}[b] \mid v_b = 1\}$
\EndWhile
\State $\mathrm{Cov}_{\mathrm{global}}
      \gets \dfrac{\sum_k \|\bm{m}_k\|_0}{\sum_k d_k}$
      \hfill$\triangleright$ $d_k$: neuron count at layer $k$
\State \Return $\mathcal{C},\; \mathrm{Cov}_{\mathrm{global}}$
\end{algorithmic}
\end{algorithm}

\section{Detailed Design of Our Approach \label{sec:methods}}
We detail the four phases shown in Figure~\ref{fig:overview} and expand Algorithm~\ref{alg:overall} to cover implementation details, including illustrative workflows and the formulation of the internal subalgorithms.

\subsection{Preprocess - Model \& Specification Synthesis}
\label{sec:synthesis}

The pipeline begins with the \textit{specification creator}, which accepts either VNNLib-compliant property files~\cite{brix2023first,2024vnncomp} (SMT-LIB format with \texttt{X\_i}/\texttt{Y\_j} variables) or PyTorch dataset/model pairs~\cite{paszke2019pytorch} to produce $B$ specification pairs $\{(\Phi^{(b)}, \Psi)\}_{b=1}^{B}$. Each input specification $\Phi^{(b)}$ constrains the feasible perturbation region around a seed, taking the form of either box constraints $\llbracket \Phi \rrbracket = \{\bm{x} \in \mathbb{R}^D \mid \lb \leq \bm{x} \leq \ub\}$ or $\ell_\infty$-balls of radius $\varepsilon$ (Eq.~\eqref{eq:inputspec}). The output specification $\Psi$ defines the property to be verified, like top-1 classification robustness (Eq.~\eqref{eq:outputspec}).

\begin{figure}[t]
    \centering
    \includegraphics[width=\linewidth]{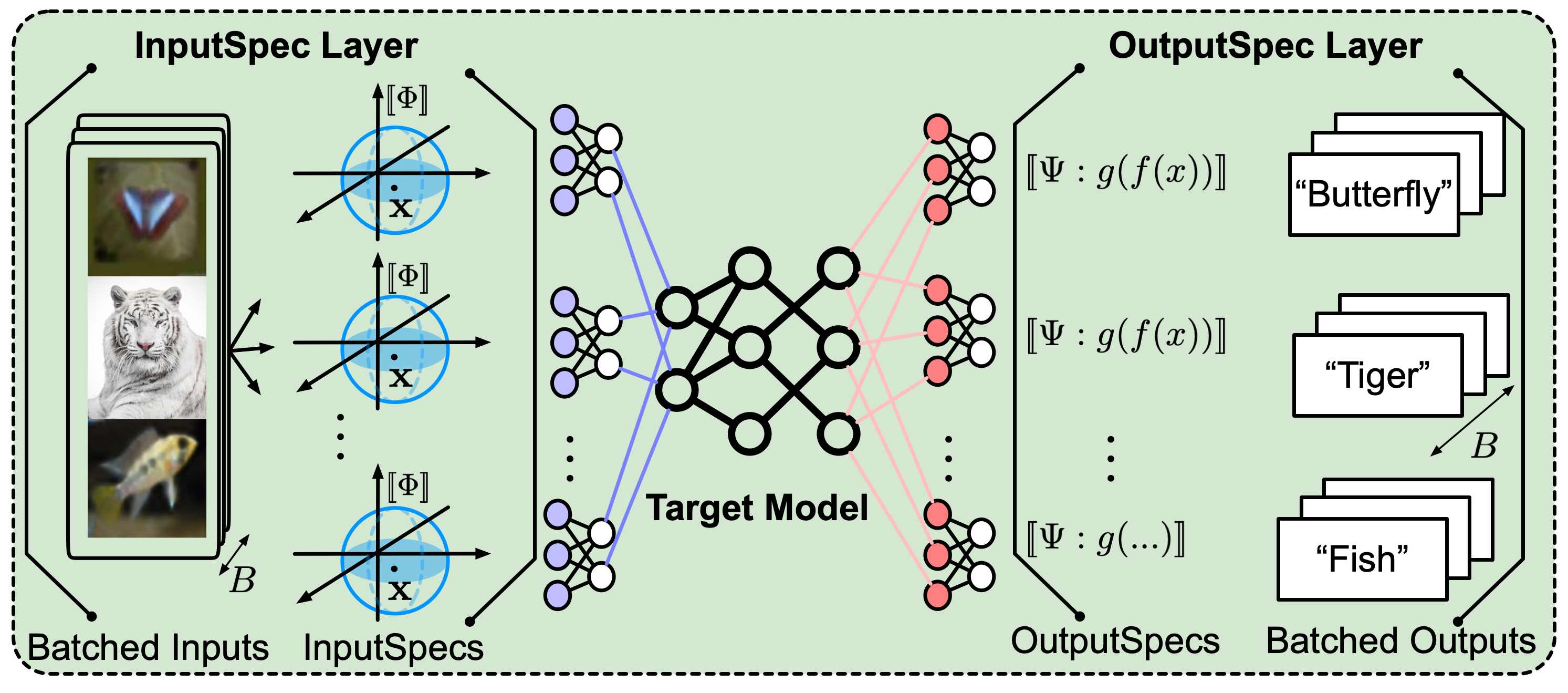}
    \caption{The \texttt{Specification Creator} accepts VNNLib property files or PyTorch dataset/model pairs to produce $B$ specification pairs $\{(\Phi^{(b)}, \Psi)\}_{b=1}^{B}$, which are batched into tensor form to construct $M_{\mathrm{wrapped}}$, (i.e., NN $f$ wrapped by non-trainable \texttt{InputSpecLayer} and \texttt{OutputSpecLayer}).}
    \label{fig:msS}
\end{figure}

As illustrated in Figure~\ref{fig:msS}, the synthesis phase merges all $B$ instances into a single batched representation by stacking along the batch dimension: seed inputs $\bm{X}_0 \in \mathbb{R}^{B \times D}$, lower bounds $\lb \in \mathbb{R}^{B \times D}$, upper bounds $\ub \in \mathbb{R}^{B \times D}$, and ground-truth labels $\bm{y} \in \{1,\dots,n'\}^B$.
The neural network $f$ is then encapsulated into a \textit{Wrapped Model} $M_{\mathrm{wrapped}}$ with the following sequential architecture, written in application order:
\begin{align}\label{eq:wrapped}
  M_{\mathrm{wrapped}}\colon\quad
  \bm{X}
  \;\xrightarrow{\texttt{InputSpecLayer}(\Phi)}\;
  f(\cdot)
  \;\xrightarrow{\texttt{OutputSpecLayer}(\Psi)}\;
  \bm{v}
\end{align}
where the arrow notation denotes the order of application: a batch of $B$ candidate inputs $\bm{X} \in \mathbb{R}^{B \times D}$ passes first through \texttt{InputSpecLayer}, which enforces per-sample feasible region constraints along the batch dimension, then through the shared target model $f$, and then through \texttt{OutputSpecLayer}, which evaluates the output specification for each of the $B$ samples paralleled to produce a batched property verdict $\bm{v} \in \{0,1\}^B$.

The \texttt{InputSpecLayer} stores the batched constraints $(\lb, \ub) \in \mathbb{R}^{B \times D}$ as non-trainable parameters. During each forward pass, it enforces the per-sample feasible region by applying the projection $\Pi_{\llbracket \Phi^{(b)} \rrbracket}$ along the batch dimension, ensuring that every candidate input passed to $f$ satisfies its corresponding input specification. The target model $f$ is shared across all $B$ samples and remains unmodified throughout the fuzzing process. The \texttt{OutputSpecLayer} evaluates the output specification $\Psi$ for each sample by computing $g(f(\bm{x}))$ as defined in Eq.~\eqref{eq:outputspec}, and returns a per-sample boolean verdict indicating whether the output property holds.

Unlike conventional software fuzzing, where each mutated input is executed as an independent sequential trial~\cite{bohme2016coverage,miller1990empirical}, the wrapped architecture makes all $B$ specifications first-class tensor dimensions. Both specification layers operate element-wise along the batch dimension, so input projection, network inference, and output property evaluation compose into a single data-parallel forward pass over all $B$ candidates simultaneously, with no external constraint lookup between steps~\cite{paszke2019pytorch,abadi2016tensorflow}. Then, every subsequent phase of the fuzzing loop inherits the batch dimension without any per-sample processing~\cite{chen2018tvm,sabne2020xla}.

\subsection{Phase 1 - Seed Selection}
\label{sec:selection}

\begin{figure}[t]
    \centering
    \includegraphics[width=\linewidth]{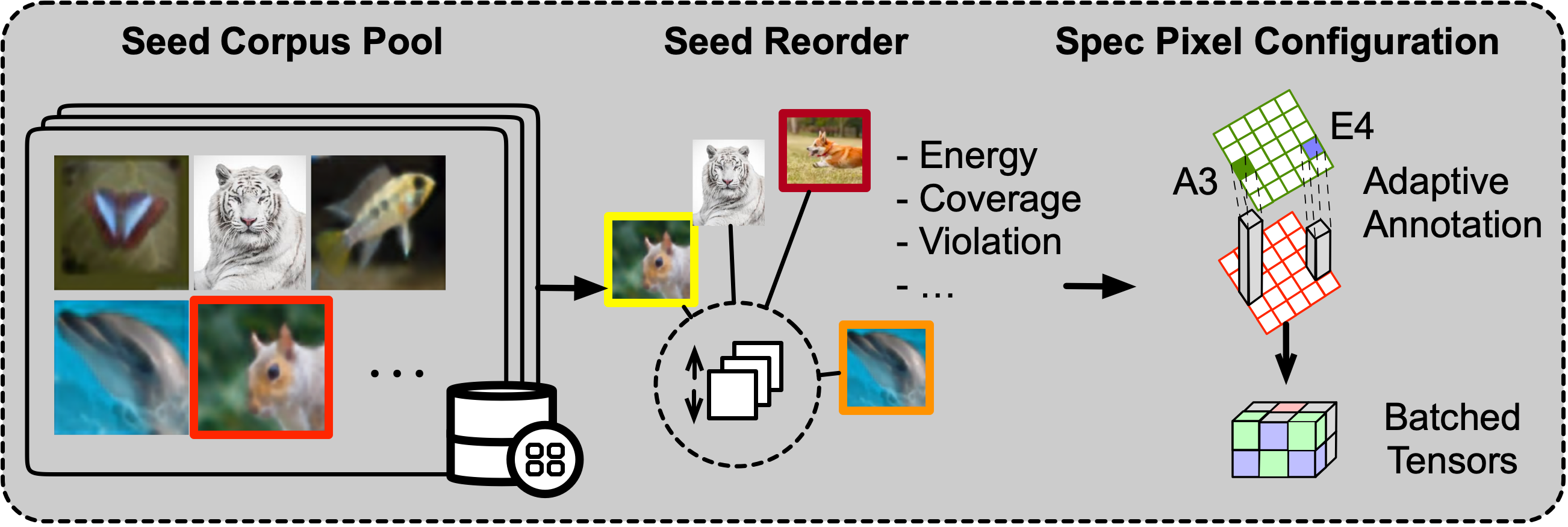}
    \caption{\textbf{Phase 1: Seed Selection.}}
    \label{fig:seed}
\end{figure}

Figure~\ref{fig:seed} shows the seed selection phase prepares specification-aware inputs for our later anisotropic mutation via a three-step pipeline:  (1) seed corpus pool management, (2) energy-based reordering which samples seeds via energy-weighted multinomial selection w.r.t high coverage contribution or prior violation history, and (3) specification pixel configuration which assigns per-dimension perturbation bounds, producing  $\bm{X}_0 \in \mathbb{R}^{B \times D}$ with lower/upper bound tensors. The output is a batched tensor representation in which every sample carries not only its input data but also per-pixel feasible bounds that guide subsequent mutation and projection operations.

\myparagraph{Seed Corpus Pool}
The synthesized seeds $\seedpool = \{\bm{x}_0^{(b)}\}_{b=1}^{B}$  are loaded into the \texttt{Seed Corpus Pool}, a tensor pool indexed along the batch dimension. Each seed stores an input tensor, an immutable original copy $\bm{x}_0^{(b)}$, a ground-truth label $y^{(b)}$ ($y^{(b)}=-1$ for unlabeled samples), an energy score $e^{(b)}$, and lineage metadata (mutation depth, parent id). Byte-level tensor hashing prevents duplicate admission. During fuzzing, mutated inputs satisfying the interestingness
criterion ($\bm{q}^{(b)}=1$ or $\bm{v}^{(b)}=1$, Section~\ref{sec:execution}) are re-inserted, progressively enriching the pool with high-coverage or violation-inducing seeds. The batch size $B$ equals the number of specification instances from the synthesis phase, guaranteeing equal fuzzing effort per specification and enabling per-sample projection via indexed access to $\lb^{(b)},\ub^{(b)}\in\mathbb{R}^D$.

\begin{algorithm}[t]
\caption{\textsc{SeedSelect}: Energy-Weighted Sampling}\label{alg:seed}
\begin{algorithmic}[1]
\Require $\seedpool = \{(\bm{x}_i,\, e_i)\}_{i=1}^{|\seedpool|}$
         \hfill$\triangleright$ $e_i \geq e_{\min} > 0$: energy updated by Alg.~\ref{alg:overall}, Line~11
\Require Batch size $B$
\Ensure  $\bm{X} \in \mathbb{R}^{B \times D}$
         \hfill$\triangleright$ batch of seeds for mutation; high-energy seeds favoured
\State $Z \gets \textstyle\sum_{j=1}^{|\seedpool|} e_j$
       \hfill$\triangleright$ normalizing constant
\If{$Z > 0$}
  \State $\bm{p} \gets \bm{e} / Z$
         \hfill$\triangleright$ energy-proportional distribution; Eq.~\eqref{eq:select}
\Else
  \State $\bm{p} \gets \bm{1} / |\seedpool|$
         \hfill$\triangleright$ uniform fallback (degenerate case)
\EndIf
\State $\mathcal{I} \sim \mathrm{Multinomial}(\bm{p},\, B,\, \mathrm{replace})$
       \hfill$\triangleright$ draw $B$ indices; allow high-energy seeds to repeat
\State $\bm{X} \gets \seedpool[\mathcal{I}].\bm{x}$
       \hfill$\triangleright$ tensor gather: $\bm{X}[b] = \bm{x}_{\mathcal{I}_b}$
\State \Return $\bm{X}$
\end{algorithmic}
\end{algorithm}

Algorithm~\ref{alg:seed} summarizes the procedure. Lines~1--4 compute the sampling distribution: normalizing constant $Z=\sum_j e_j$ converts energies to probabilities, with uniform fallback when $Z=0$ (Eq.~\eqref{eq:select}). Line~5 draws $B$ indices with replacement (high-energy seeds may repeat); Line~6 gathers the result into the tensor for mutation.

\myparagraph{Energy-Based Seed Reordering} At each iteration, the corpus $\seedpool$ is sampled via energy-weighted multinomial selection with replacement:
\begin{align}\label{eq:select}
  P(\text{select seed } b)
    = e_b \Big/ \sum_{j=1}^{|\seedpool|} e_j
\end{align}
where $e_b$ is the energy of seed $b$. Seeds that contributed to coverage growth or triggered property violations accumulate higher energy (Section~\ref{sec:execution}), so that subsequent iterations concentrate effort on the most promising regions of the
input space.

\myparagraph{Specification Pixel Configuration}
Rather than treating all dimensions uniformly as in
Eq.~\eqref{eq:inputspec}~\cite{tishby2015deep}, the specification assigns each
dimension $d$ an individual spec radius, giving a per-dimension spec radius
vector $\bm{\varepsilon}^{(b)}\in\mathbb{R}^D$ and feasible region:
\begin{align}\label{eq:pixel_spec}
  \llbracket\Phi^{(b)}_{\mathrm{px}}\rrbracket
    = \bigl\{\bm{x}\in\mathbb{R}^D \mid
      |\bm{x}_d - \bm{x}_{0,d}^{(b)}| \leq \varepsilon_d^{(b)},\;\forall\,d\bigr\}
\end{align}
$\varepsilon_d^{(b)}$ is sourced from VNNLib per-dimension bounds
$(\lb^{(b)},\ub^{(b)})$, user-defined spatial budgets, or a single uniform
$\varepsilon$ (recovering the standard $\ell_\infty$-ball). Seeds are then
stacked to form $\bm{X}_0\in\mathbb{R}^{B\times D}$ with bounds
$\lb,\ub\in\mathbb{R}^{B\times D}$:
\begin{align}\label{eq:pixel_bounds}
  l_{b,d} = \bm{x}_{0,d}^{(b)} - \varepsilon_d^{(b)}, \qquad
  u_{b,d} = \bm{x}_{0,d}^{(b)} + \varepsilon_d^{(b)}
\end{align}

\myparagraph{Batched Tensors}
These batched bounds are configured into the \texttt{InputSpecLayer} of
$M_{\mathrm{wrapped}}$ (Section~\ref{sec:synthesis}), which enforces
per-sample, per-dimension projection during every forward inference. 
When
$\varepsilon_d^{(b)} = \varepsilon$ for all $d$, the formulation reduces to the standard uniform $\ell_\infty$-ball, maintaining backwards compatibility with conventional robustness specifications.
This non-uniform bound structure creates a direct link to the downstream mutation phase (Section~\ref{sec:mutation}). Each dimension's feasible range $r_d^{(b)} = u_{b,d} - l_{b,d} = 2\,\varepsilon_d^{(b)}$ varies across dimensions: wider bounds admit larger perturbations while tight bounds enforce finer exploration. The mutation engine can later derive its step size $\eta$ from these per-dimension bounds via a single scale factor $s \in (0,1)$, automatically calibrating mutation granularity to the specification structure
without manual tuning.

\begin{figure}[b]
    \centering
    \includegraphics[width=\linewidth]{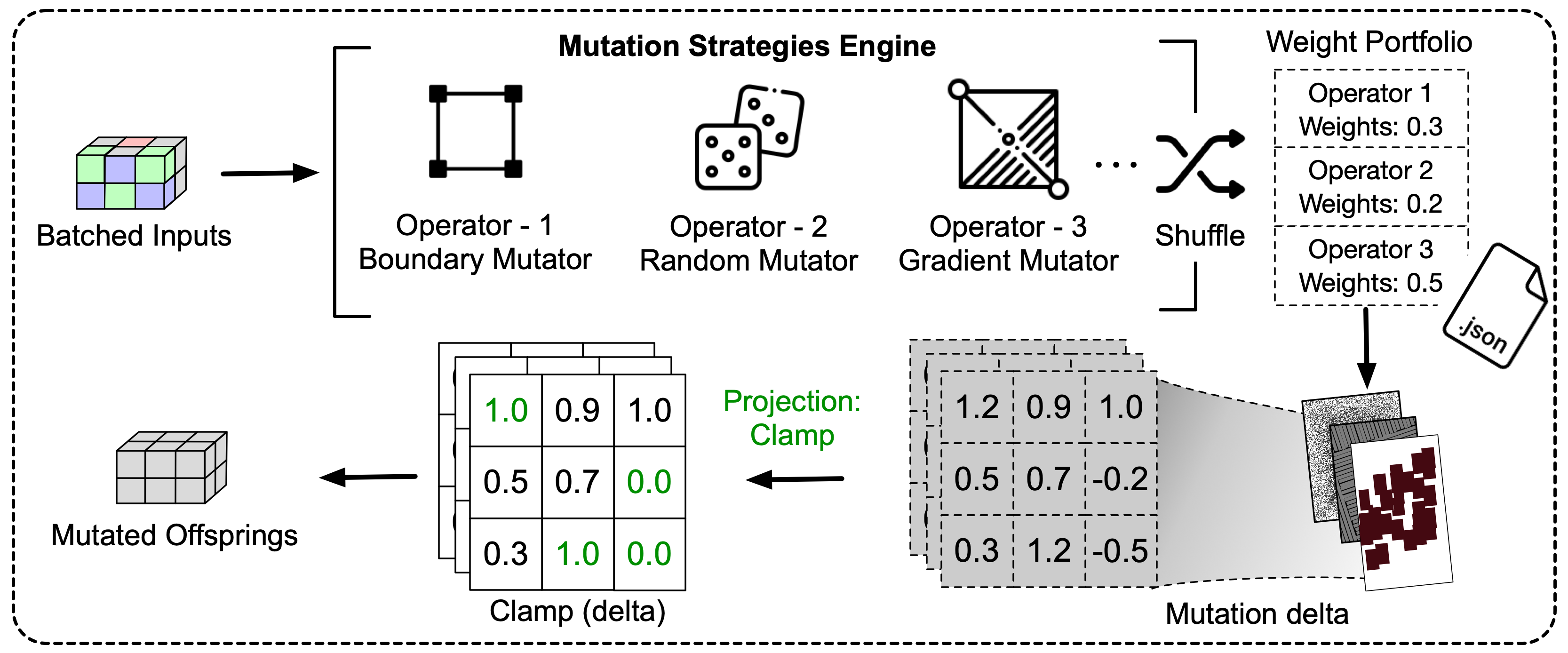}
    \caption{Phase 2: Mutation.}
    \label{fig:mutate}
\end{figure}

\subsection{Phase 2 - Mutation}
\label{sec:mutation}

\begin{algorithm}[t]
\caption{\textsc{Mutate}: Batch Mutation with Projection}\label{alg:mutate}
\begin{algorithmic}[1]
\Require $\bm{X} \in \mathbb{R}^{B \times D}$
         \hfill$\triangleright$ current seed batch from \textsc{SeedSelect}
\Require $(\lb,\ub) \in \mathbb{R}^{B \times D}$
         \hfill$\triangleright$ per-sample, per-dimension spec bounds
\Require Step size $\eta$ (Alg.~\ref{alg:overall}, Line~6);\;
         strategy weights $W$;\; PGD steps $T$;\;
         $M_{\mathrm{wrapped}}$ (gradient strategies only)
\Ensure  $\tilde{\bm{X}} \in \mathbb{R}^{B \times D}$
         \hfill$\triangleright$ $\tilde{\bm{X}}[b] \in \llbracket\Phi^{(b)}\rrbracket$ for all $b$
\State $\mu \sim \mathrm{Categorical}\bigl(W / \textstyle\sum_\nu w_\nu\bigr)$
       \hfill$\triangleright$ Eq.~\eqref{eq:strategy_select}
\If{$\mu = \textsc{Gradient}$}
       \hfill$\triangleright$ $T{=}1$: \textsc{fgsm};\; $T{>}1$: \textsc{pgd}
  \State $\tilde{\bm{X}} \gets \bm{X} + \mathcal{U}[-\eta,\,\eta]$
         \hfill$\triangleright$ random initialization; broadcasts if $\eta \in \mathbb{R}$
  \For{$t = 0, \ldots, T{-}1$}
    \State $\tilde{\bm{X}} \!\gets\!
           \Pi_{\llbracket\Phi\rrbracket}\!\bigl(
           \tilde{\bm{X}} \!+\! \tfrac{2}{T}\eta \odot
           \mathrm{sign}(\nabla_{\tilde{\bm{X}}}\,
           \mathcal{L}(M_{\mathrm{wrapped}}(\tilde{\bm{X}})))\bigr)$
           \hfill$\triangleright$ element-wise step; Eq.~\eqref{eq:pgd_iter}
  \EndFor
\ElsIf{$\mu = \textsc{Boundary}$}
  \State $\tilde{\bm{X}} \gets \bm{X}
         + \tfrac{1}{2}\eta \odot \mathrm{sign}(\bm{r})$,\;
         $\bm{r} \sim \mathcal{N}(\bm{0},\bm{I})$
\ElsIf{$\mu = \textsc{Random}$}
  \State $\tilde{\bm{X}} \gets \bm{X}
         + \mathcal{N}\!\bigl(\bm{0},\,(\tfrac{1}{2}\eta)^2\bm{I}\bigr)$
\EndIf
\State $\tilde{\bm{X}} \gets \Pi_{\llbracket\Phi\rrbracket}(\tilde{\bm{X}})$
       \hfill$\triangleright$ Eq.~\eqref{eq:proj_box}/\eqref{eq:proj_linf};\;
       enforces $\lb \leq \tilde{\bm{X}} \leq \ub$
\State \Return $\tilde{\bm{X}}$
\end{algorithmic}
\end{algorithm}

Figure~\ref{fig:mutate} shows the workflow of our mutation phase, which selects a single operator from a weighted portfolio (\textsc{Boundary}, \textsc{Random}, or \textsc{Gradient}) and applies it uniformly to the entire batch of $B$ inputs, producing a mutation delta tensor. The resulting candidates are then projected back onto the feasible regions via clamp, yielding the mutated batch tensor.

Algorithm~\ref{alg:mutate} formalises the batch mutation procedure. Line~1
samples a single strategy $\mu$ from the weighted portfolio
(Eq.~\eqref{eq:strategy_select}). Lines~2--3 implement the gradient operator:
$\gamma$ is set proportional to $\eta$ and $T$ projected gradient-ascent steps
are applied (Eq.~\eqref{eq:pgd_iter}); $T{=}1$ recovers \textsc{fgsm}
(Eq.~\eqref{eq:fgsm}). Lines~4 and~5 implement the boundary and random
operators. Line~6 applies hard constraint projection
(Eq.~\eqref{eq:proj_box}/\eqref{eq:proj_linf}), ensuring every output
candidate lies within its feasible region.

\myparagraph{Strategy Portfolio}
A single strategy is applied to the entire batch, keeping forward and backward
passes fully vectorised:
\begin{align}\label{eq:strategy_select}
  P(\text{select } \mu) = w_\mu \big/ \textstyle\sum_{\nu} w_\nu
\end{align}
where $w_\mu$ is the configurable weight of strategy $\mu$.

\begin{table}[!t]
\caption{Mutation Strategies (selected via Line 1 Algorithm~\ref{alg:mutate})}\label{tab:strategies}
\centering

\resizebox{0.98\linewidth}{!}{
\begin{tabular}{llcl}
\toprule
\textbf{Strategy} & \textbf{Type} & \textbf{Weight} & \textbf{Purpose} \\
\midrule
\textsc{Gradient} & Gradient & 0.5 &
  Adversarial search~\cite{madry2017towards,goodfellow2015explaining} \\
\textsc{Boundary} & Heuristic & 0.2 &
  Push toward specification boundary~\cite{carlini2017towards} \\
\textsc{Random}   & Heuristic & 0.3 &
  Isotropic Gaussian exploration~\cite{odena2019tensorfuzz} \\
\bottomrule
\end{tabular}%
}
\end{table}

\myparagraph{Gradient-Driven Operator}
We implement \textsc{pgd}~\cite{madry2017towards} and its single-step special
case \textsc{fgsm}~\cite{goodfellow2015explaining} ($T{=}1$). Starting from a
random point in $[\bm{x}^{(b)} - \eta,\, \bm{x}^{(b)} + \eta]$, $T$
projected gradient-ascent steps are applied:
\begin{align}\label{eq:pgd_iter}
\bm{x}_{t+1} = \Pi_{\llbracket \Phi^{(b)} \rrbracket}\!\left(\bm{x}_t + \tfrac{2}{T}\,\eta \odot \operatorname{sign}\!\big(\nabla_{\bm{x}} \mathcal{L}(\bm{x}_t)\big)\right), \quad t = 0, \ldots, T-1
\end{align}
where $\eta$ is a broadcastable perturbation scale: in isotropic mode,
$\eta \in \mathbb{R}$ is a scalar, while in anisotropic mode,
$\eta = S \in \mathbb{R}^{B \times D}$ provides element-wise step sizes.
Here $\Pi_{\llbracket\Phi^{(b)}\rrbracket}$ projects onto the spec-feasible region,
and $\mathcal{L}$ is the adversarial objective (cross-entropy when a label
$y^{(b)}$ is available, output variance otherwise). When $T{=}1$,
Eq.~\eqref{eq:pgd_iter} reduces to \textsc{fgsm}:
\begin{align}\label{eq:fgsm}
\tilde{\bm{x}}^{(b)}
= \bm{x}^{(b)}
+ \eta \odot \operatorname{sign}\!\bigl(
\nabla_{\bm{x}}\,\mathcal{L}(\bm{x})
\bigr)\big|_{\bm{x}=\bm{x}^{(b)}}
\end{align}
where $\odot$ reduces to scalar multiplication in the isotropic case.
Both variants operate over all $B$ samples in a single backward pass.

\myparagraph{Heuristic Operators}
Both operators require no gradient and incur negligible overhead. The
\textsc{Boundary} operator pushes seeds toward the extremes of the feasible
region: $\tilde{\bm{X}} = \bm{X} + \tfrac{1}{2}\eta \odot \mathrm{sign}(\bm{r})$,
$\bm{r}\sim\mathcal{N}(\bm{0},\bm{I})$. 
Here $\odot$ is element-wise multiplication, broadcasting when $\eta \in \mathbb{R}$ and applying dimension-wise scaling when $\eta = \mathcal{S} \in \mathbb{R}^{B \times D}$.
The \textsc{Random} operator applies
isotropic Gaussian noise: $\tilde{\bm{X}} = \bm{X} +
\mathcal{N}(\bm{0},(\tfrac{1}{2}\eta)^2\bm{I})$. Both use a halved magnitude
$\tfrac{1}{2}\eta$ for conservative exploration near the current seed.

\myparagraph{Adaptive Perturbation Sizing}
The step size $\eta$ is derived from the specification bounds under two modes.
In \textbf{fixed mode}, $\eta$ is a user-specified constant (default $0.01$),
serving as a scale-agnostic baseline. In \textbf{adaptive mode}, $\eta$ is
computed from the per-dimension feasible range $r_{b,d} = \ub_{b,d} -
\lb_{b,d}$ scaled by $s \in (0,1]$. The \emph{isotropic} variant yields a
single scalar from the mean range:
\begin{align}\label{eq:perturb_scalar}
  \eta = s \cdot \frac{1}{BD}\sum_{b,d} r_{b,d}
\end{align}
The \emph{anisotropic} variant assigns a separate step size per dimension:
\begin{align}\label{eq:perturb_perdim}
  \mathcal{S}_{b,d} = s \cdot r_{b,d}
    = s \cdot (\ub_{b,d} - \lb_{b,d})
\end{align}
so dimensions with wider bounds receive proportionally larger steps. The
factor $s$ has a clean traversal interpretation: $1/s$ steps are needed to
traverse any dimension's full feasible range adaptively. The \textsc{pgd} step size
$\gamma = 2\,\mathrm{mean}(\eta)/T$ is derived from $\eta$, so both the
local search scope and the gradient step size scale automatically with the
specification. \revision{Adaptive scaling refines search granularity in a specification-aware manner; throughput gains derive primarily from batching, while anisotropic scaling improves exploration on heterogeneous specifications.}

\myparagraph{Hard Constraint Projection}
After mutation, each candidate is projected onto the feasible region. For
\textbf{box constraints}, the projection is an element-wise clamp:
\begin{align}\label{eq:proj_box}
  \tilde{\bm{x}}_{b,d}
    = \mathrm{clamp}\bigl(\tilde{\bm{x}}_{b,d},\; \lb_{b,d},\; \ub_{b,d}\bigr)
\end{align}
For $\ell_\infty$-ball constraints, the perturbation is clamped relative to the original input $\bm{x}_0^{(b)}$, stored immutably in the
corpus:
\begin{align}\label{eq:proj_linf}
  \tilde{\bm{x}}^{(b)}
    = \bm{x}_0^{(b)}
      + \mathrm{clamp}\bigl(
          \tilde{\bm{x}}^{(b)} - \bm{x}_0^{(b)},\; -\varepsilon,\; \varepsilon
        \bigr)
\end{align}
Anchoring to the original input preserves the $\ell_\infty$ invariant across
mutation chains of arbitrary depth. Both projections are fully vectorized over
all $B$ samples.

\subsection{Phases 3 and 4 - Execution \& Feedback}
\label{sec:execution}

\begin{figure}[!t]
    \centering
    \includegraphics[width=0.85\linewidth]{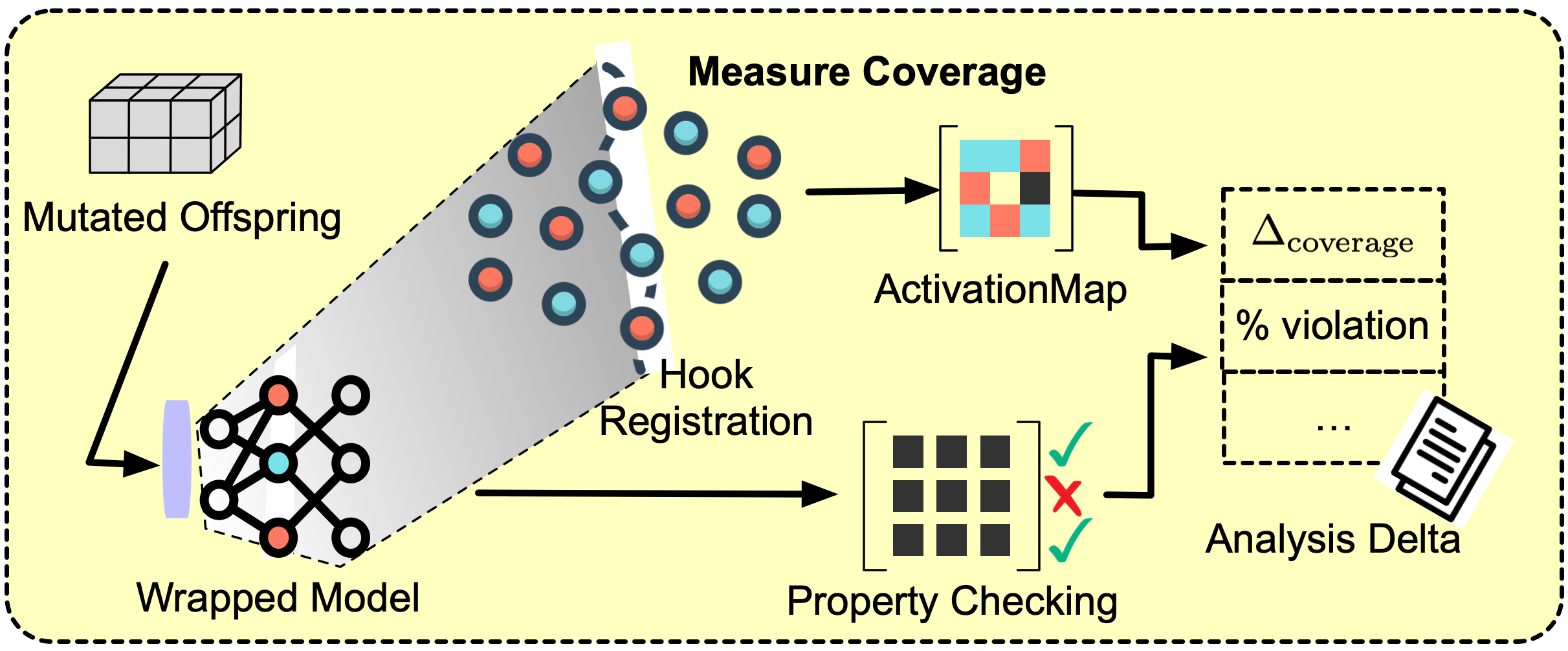}
    \caption{Phase 3: Execution.}
    \label{fig:hook}
\end{figure}

Figure~\ref{fig:hook} shows that the projected candidate batch is passed through $M_{\mathrm{wrapped}}$ in a single batched forward pass. PyTorch forward hooks registered on computational layers (ReLU, Linear, Conv2d) intercept intermediate activations to populate the activation map $A$.

\begin{algorithm}[t]
\caption{\textsc{ExecFeedback}: Inference \& Feedback}\label{alg:exec}
\begin{algorithmic}[1]
\Require $\tilde{\bm{X}} \in \mathbb{R}^{B \times D}$,\; $M_{\mathrm{wrapped}}$
         \hfill$\triangleright$ mutated batch from Alg.~\ref{alg:mutate}
\Require Coverage $\{\bm{m}_k\}_{k=1}^{K}$;\;
         $\tau$,\; $\alpha$,\; $\beta$,\; $e_{\min}$
         \hfill$\triangleright$ shared with Alg.~\ref{alg:overall}
\Ensure  $\bm{q},\bm{v} \in \{0,1\}^B$;\; $\bm{e} \in \mathbb{R}^B$;\; $\mathcal{C}$
\State $(\hat{\bm{Y}},\, A) \gets f(\tilde{\bm{X}})$
       \hfill$\triangleright$ single batched forward pass;\;
       $A = \{k \mapsto \bm{a}_k \in \mathbb{R}^{B \times d_k}\}$ via hook instrumentations
\State $\bm{q} \gets \bm{0}^B$
\For{each hooked layer $k = 1,\ldots,K$}
  \State $\bm{F}_k \gets \bigl[|\bm{a}_{k,j}^{(b)}| > \tau\bigr]_{b,j}
         \in \{0,1\}^{B \times d_k}$
         \hfill$\triangleright$ per-sample firing matrix;\; Eq.~\eqref{eq:neuron_fired}
  \State $\bm{n}_k \gets
         \bigl(\textstyle\bigvee_{b}\,\bm{F}_k[b,\cdot]\bigr)
         \wedge \lnot\,\bm{m}_k$
         \hfill$\triangleright$ newly covered neurons at layer $k$
  \State $\bm{q} \gets \bm{q} \lor [\bm{F}_k\,\bm{n}_k > \bm{0}]$;\quad
         $\bm{m}_k \gets \bm{m}_k \lor \textstyle\bigvee_{b}\,\bm{F}_k[b,\cdot]$
         \hfill$\triangleright$ interestingness mask and coverage state (in-place)
\EndFor
\State $\bm{v} \gets [\Psi(\hat{\bm{Y}}) = \mathit{false}]$
       \hfill$\triangleright$ per-sample violation  (Alg.~\ref{alg:overall}, Line~9)
\State $\bm{e} \gets \max(\bm{q}\alpha + \bm{v}\beta,\; e_{\min})$
       \hfill$\triangleright$ energy scores (Alg.~\ref{alg:overall}, Line~11)
\State $\mathcal{C} \gets \{\tilde{\bm{X}}[b] \mid v_b = 1\}$
       \hfill$\triangleright$ counterexample inputs
\State \Return $\bm{q},\; \bm{v},\; \bm{e},\; \mathcal{C}$
\end{algorithmic}
\end{algorithm}

Algorithm~\ref{alg:exec} expands Phases~4--5 of Alg.~\ref{alg:overall}.
Line~1 executes a single batched forward pass and captures intermediate
activations via hooks. Lines~3--8 iterate over $K$ hooked layers: Line~4
computes the per-sample firing matrix (Eq.~\eqref{eq:neuron_fired}), Line~5
identifies newly covered neurons, and Line~6 updates the interestingness mask
$\bm{q}$ and coverage state $\bm{m}_k$ in-place. Lines~9--10 evaluate
violations and compute energy scores (Alg.~\ref{alg:overall}, Lines~9--11), and
Line~10 collects counterexample inputs.

\myparagraph{Batched Inference and Activation Capture}
The projected batch is passed through $M_{\mathrm{wrapped}}$ in a single forward pass, yielding $\hat{\bm{Y}} \in \mathbb{R}^{B \times D'}$ and activation map $A = \{k \mapsto \bm{a}_k \in \mathbb{R}^{B \times
d_k}\}_{k=1}^{K}$, where PyTorch forward hooks~\cite{paszke2019pytorch}
intercept intermediate activations.
For convolutional layers with activations in $\mathbb{R}^{B \times C_k \times
H \times W}$, spatial dimensions are reduced to per-channel scalars via absolute-value max-pooling; fully connected and ReLU layers require no reduction.

\myparagraph{Neuron Coverage Tracking}
Neuron $(k,j)$ is considered fired by sample $b$ if its activation exceeds
threshold $\tau$:

\begin{align}\label{eq:neuron_fired}
  F_{k,j}^{(b)} = \bigl[|\bm{a}_{k,j}^{(b)}| > \tau\bigr]
\end{align}
Two strategies interpret this signal. The \emph{global union} strategy
(Lines~5--6) maintains a persistent mask $\bm{m}_k$ per layer, updated
monotonically via bitwise OR; a sample is interesting if it fires at least
one previously uncovered neuron, analogous to the new-edge criterion in
coverage-guided fuzzers~\cite{xie2019deephunter}. Global coverage is tracked
as $\mathrm{Cov}_{\mathrm{global}} = \sum_k\|\bm{m}_k\|_0 / \sum_k d_k$
(Alg.~\ref{alg:overall}, Line~14). The \emph{best-input} strategy replaces
the persistent mask with a running maximum coverage scalar $c_{\max}$; a sample is interesting if its individual coverage ratio exceeds $c_{\max}$.

\myparagraph{Property Checking}
The \texttt{OutputSpecLayer} evaluates $\Psi$ for all $B$ samples during the
forward pass (Section~\ref{sec:synthesis}), producing the violation mask
$\bm{v}$ (Alg.~\ref{alg:overall}, Line~9) with no additional model evaluation. \revision{The batched property-evaluation formulation expresses margin, range, and linear-inequality properties through a single dispatch; in our experiments the evaluated property is top-1 (classification) robustness, flagging $v_b = 1$ when $\arg\max_c\,\hat{y}_{b,c} \neq y^{(b)}$.}

\myparagraph{Energy-Based Feedback}
Per-sample energies $\bm{e} = \max(\bm{q}\alpha + \bm{v}\beta,\,e_{\min})$
(Alg.~\ref{alg:overall}, Line~11) weight violations ($\beta=100$) an order of
magnitude higher than coverage gains ($\alpha=10$), reflecting the primary
objective of counterexample discovery; the minimum clamp $e_{\min}=0.1$
prevents seed starvation. A candidate is re-inserted when $q_b \lor v_b$
(Alg.~\ref{alg:overall}, Line~12), with a hash preventing redundant entries.

\section{Experiments}
\myparagraph{Implementation} We have conducted extensive experiments to evaluate the performance of
our approach under both isotropic (\iso) and anisotropic (\ani) settings. The approach is implemented in Python~3.12. All experiments are conducted on a machine running Ubuntu~24.04.4~LTS, equipped with an NVIDIA RTX PRO 6000 Blackwell Max-Q GPU (${\approx}$96\,GB VRAM), an Intel Core Ultra~7 265K CPU (20 cores, up to 6.5\,GHz), and 128\,GiB of memory. 
%
\revision{Our latest implementation is available in the  ACT platform at \textbf{\url{https://github.com/SVF-tools/ACT}.}}

\begin{table}[htbp]
\centering
\caption{Benchmark Summarization. $G$ is the number of model groups, each pairing a fixed model with its specification instances. Batch size is the number of (spec, input) pairs per group, i.e., the maximum $B$ processed in a batch.}

\label{tab:benchmarks}

\resizebox{0.98\linewidth}{!}{%
\setlength{\tabcolsep}{3pt}
\begin{tabular}{l l c r c c}
\toprule
\textbf{Benchmark} & \textbf{Model(s)} & \textbf{Input Shape} & \textbf{\#Params} & $G$ & \textbf{Batch Size} $B$\\
\midrule
  \textsc{TrafficSigns}
  & 3 QCNNs
  & $3{\times}\{30, 48, 64\}^2$
  & 0.9--1.8\,M
  & 3
  & 14--15 \\\midrule
\textsc{Cifar100}
  & ResNet-medium
  & \multirow{2}{*}{$3{\times}32{\times}32$}
  & 2.54\,M
  & \multirow{2}{*}{2}
  & 99 \\
  & ResNet-large
  &
  & 3.81\,M
  &
  & 100 \\\midrule
\textsc{TinyImageNet}
  & ResNet
  & $3{\times}64{\times}64$
  & 3.62\,M
  & 1
  & 199 \\

\bottomrule
\end{tabular}%
}
\end{table}

\myparagraph{Benchmarks} We evaluate on three benchmark categories drawn from VNN-COMP~\cite{2024vnncomp} and standard PyTorch model suites~\cite{brix2023first,2024vnncomp} in Table~\ref{tab:benchmarks}. 
Our fuzzer's main objective is to generate inputs that serve as concrete counterexamples to the output property $\Psi$. Since our batch fuzzing is model-centric, all $B$ specification instances within a batch must share the same network with identical weights, where $B$ denotes the batch size and each instance corresponds to an input together with its associated input constraints and output property.
To accommodate this requirement, we partition each benchmark into $G$ \emph{model groups}, where each group consists of a fixed model paired with multiple specification instances. Accordingly, the effective batch size $B$ is defined per model group as the number of specification instances associated with that model, rather than the total number of instances across all $G$ groups. We list the benchmarks used in our evaluation below.
\begin{itemize}
\item \textsc{TrafficSigns}~\cite{stallkamp2012man} provides three quantised CNNs
at resolutions $30{\times}30$, $48{\times}48$, and $64{\times}64$
(${\approx}$0.9--1.8\,M parameters) for 43-class traffic sign recognition, forming three
model groups of $B{=}14$ or $15$.
\item \textsc{Cifar100}~\cite{krizhevsky2009learning} uses two ResNet variants on
$3{\times}32{\times}32$ images: ResNet-medium (${\approx}$2.54\,M parameters,
$B{=}99$) and ResNet-large (${\approx}$3.81\,M parameters, $B{=}100$).
\item \textsc{TinyImageNet}~\cite{le2015tiny} evaluates a single ResNet
(${\approx}$3.62\,M parameters) on all 199 robustness specifications over
$3{\times}64{\times}64$ images, giving the largest batch size $B{=}199$.
\end{itemize}

\myparagraph{Evaluation Metrics}
The effectiveness is evaluated via below:
\begin{itemize}
    \item \emph{Violation count} records the total number of specification-violating
counterexamples discovered, reflecting the primary objective of the fuzzing
campaign.
\item \emph{Time to first violation} (TTFV) captures the wall-clock time elapsed
before the first counterexample is found, providing a latency-oriented measure
of how quickly a configuration can expose property violations.
\item \emph{Throughput} (Thpt), defined as the number of candidate instances (mutated cases) generated per second, quantifies the raw testing efficiency afforded by batched execution.
\end{itemize}

\myparagraph{Experimental Settings}
We compare against a controlled baseline that captures the canonical one-input-per-iteration fuzzing paradigm underlying DeepHunter~\cite{xie2019deephunter}, which does not have an official open-source implementation. We implement this baseline within our framework (where $B{=}1$ with fixed scalar $\eta$, as in Table~\ref{tab:fuzzing-comparison}), enabling a fair comparison that isolates the effects of tensor-level parallelism and adaptive perturbation scaling. All other parameters are held constant across configurations: coverage criterion~\cite{pei2017deepxplore} with $\tau{=}0.1$, energy constants ($\alpha{=}10$, $\beta{=}100$, $e_{\min}{=}0.1$), and per-instance timeout $t_{\max}{=}60$s. We do not compare directly against TensorFuzz~\cite{odena2019tensorfuzz}, which also follows the one-input-per-iteration paradigm: although it batches mutated inputs for model evaluation, its mutation is NumPy-based, and coverage feedback iterates per element in Python, rather than operating as tensor-based transformations (e.g., PyTorch tensors). Its implementation~\cite{tensorfuzz_github} has been unmaintained since 2019, and incompatible input specifications limit its use on our large-scale test suites.

We primarily study the following research questions (\textbf{RQs}):
\begin{itemize}
    \item \textbf{RQ1 } What is speedup and throughput 
of our batch, fuzzing over the sequential baseline method? 
    \item  \textbf{RQ2  } To what extent does the batched approach, including both isotropic and anisotropic modes, account for the performance gains over the sequential baseline?
    \item \textbf{RQ3  } To what extent does anisotropic perturbation sizing account for the performance gains w.r.t scale factor $s$?
    \item \textbf{RQ4 } To what extent does the tensor batch size $B$ have a statistically significant effect on violation yield?
\end{itemize}

\subsection{Fuzzing Methods Comparison (RQ1 \& RQ2)}

We examine whether tensor-level execution yields measurable throughput gains that achieve higher violation counts within a fixed budget. 
As illustrated in Figure~\ref{fig:overview}, the wrapped model
$M_{\mathrm{wrapped}}$ shares a single network $f$ across all $B$
specification instances, amortizing the forward pass cost over the entire
batch. All violation counts are aggregated across all model groups $G$. 

\begin{table}[htpb]
\centering
\caption{Fuzzing results across three benchmark categories.}
\label{tab:results}
\renewcommand{\arraystretch}{1.3}
\resizebox{0.98\linewidth}{!}{%
\begin{tabular}{l l r r r r}
\toprule
\textbf{Benchmark} & \textbf{Method}
& \makecell[c]{\textbf{Violations}\\\textbf{(\#)}}
& \makecell[c]{\textbf{TTFV}\\\textbf{(sec)}}
& \makecell[c]{\textbf{Neuron}\\\textbf{Cov.\,(\%)}}
& \makecell[c]{\textbf{Thpt.}\\\textbf{(it/s)}} \\
  \midrule
\textsc{TrafficSigns}
  & \fix           & 19229 & 60.12 & 100\%& 79.6 \\
  & \iso          & 73316 & 0.06 & 100\%&407.3 \\
  & \ani                   & 75208 & 0.06 & 100\%&417.9 \\
\midrule
\textsc{Cifar100}
  & \fix           & 14562 & 60.14 & 57.19\%&15.2 \\
  & \iso          & 60242 & 0.60 & 64.06\%&516.4 \\
  & \ani           & 61527 & 0.60 &64.07\%& 524.0 \\
\midrule
\textsc{TinyImageNet}
  & \fix           & 16890 & 60.18 &59.04\%& 10.9 \\
  & \iso          & 23125 & 1.20 & 70.04\%&395.1 \\
  & \ani                   & 25274 & 1.09 & 70.04\%&430.7 \\
\bottomrule
\end{tabular}}
\end{table}

As shown in Table~\ref{tab:results}, both \iso and \ani configurations achieve up to 40$\times$ throughput over \fix, with gains of 34--40$\times$ on \textsc{Cifar100} and \textsc{TinyImageNet} and 5$\times$ on \textsc{TrafficSigns}, where the lighter network architecture allows \fix to run faster (79.6\,instances/s), leaving less relative room for batch speedup. Notably, batch execution processes specifications in a single 60\,s window, whereas \fix allocates 60\,s per specification instance sequentially ($\#\text{instances} \times 60$\,s in total); despite this extended budget, \fix accumulates only 14K--19K violations against 60K--75K for batch execution.

The TTFV of \fix being close to the per-instance timeout suggests a throughput bottleneck rather than an inability to discover violations, as Table~\ref{tab:batch-advantage} shows it continues accumulating violations beyond 60\,s. Neuron Coverage (Neuron Cov.) is reported with a fixed threshold $\tau = 0.1$ across all configurations; As neuron coverage is a coarse structural metric~\cite{odena2019tensorfuzz} that does not necessarily correlate with violation discovery, absolute values should be interpreted accordingly.
On \textsc{TrafficSigns}, all methods reach full neuron coverage, yet batch execution still finds $4\times$ more violations, indicating gains are primarily due to higher throughput. On \textsc{Cifar100} and \textsc{TinyImageNet}, batch methods also achieve higher coverage (64\% vs.\ 57\% and 70\% vs.\ 59\%) within the same 60\,s budget, suggesting improved exploration efficiency per unit time.

\myparagraph{RQ1: Batch Speedups} Tensor-level parallelism yields throughput gains that scale with 
batch size: up to 40$\times$ on \textsc{Cifar100} and \textsc{TinyImageNet} (where $B \geq 99$), and 5$\times$ on \textsc{TrafficSigns} ($B{=}14$--$15$). In all cases, batch execution completes the full specification campaign in a single 60\,s window, exposing the first counterexample within 1.2\,s versus the sequential baseline which requires up to $B \times 60$\,s to process all specifications.

\begin{figure}[!t]
\centering
\begin{subfigure}[b]{0.46\linewidth}
  \includegraphics[width=\linewidth]{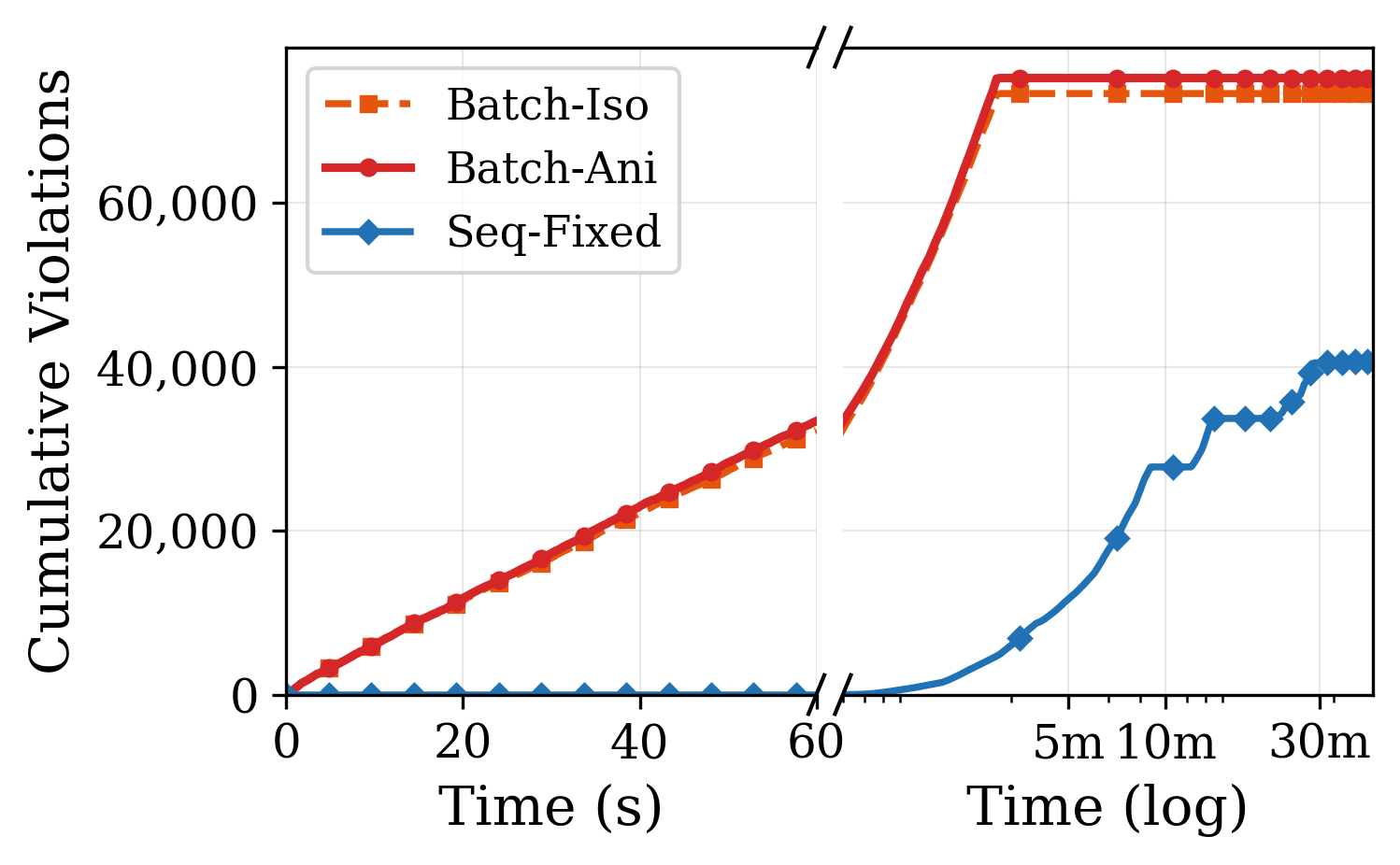}

  \caption{TrafficSigns} \label{fig:cu-traffics}
\end{subfigure}
\begin{subfigure}[b]{0.46\linewidth}
  \includegraphics[width=\linewidth]{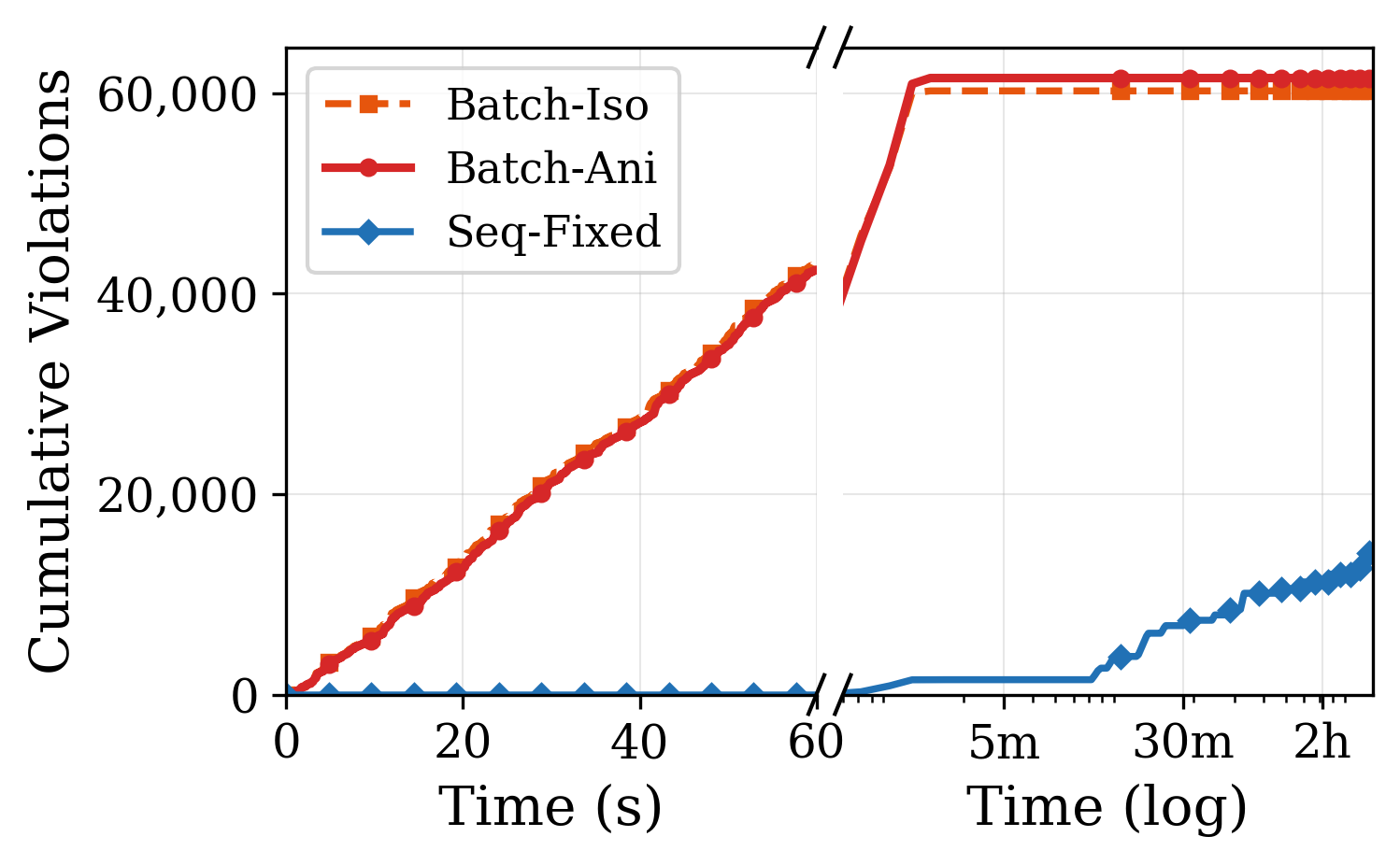}

  \caption{CIFAR-100}\label{fig:cu-cifar}
\end{subfigure}
\begin{subfigure}[b]{0.46\linewidth}
  \includegraphics[width=\linewidth]{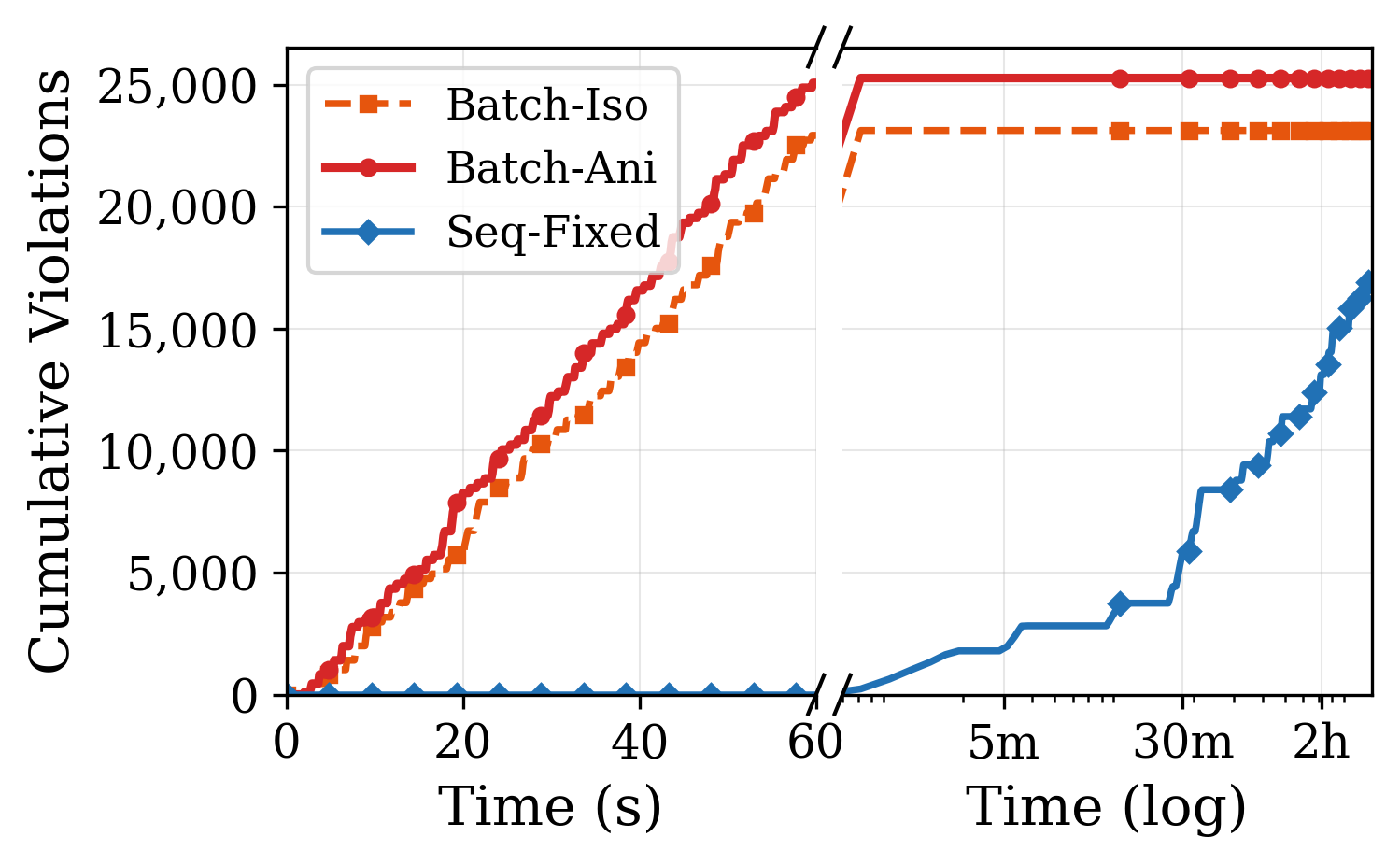}

  \caption{TinyImageNet}\label{fig:cu-imagenet}
\end{subfigure}

\caption{Cumulative violations over three benchmarks.}
\label{fig:violtime}
\end{figure}
  
  

\myparagraph{RQ2: Cumulative Violations} We further examine whether the throughput advantage translates into a sustained violation discovery advantage over the full fuzzing runs.

Figure~\ref{fig:violtime} plots cumulative violations over wall-clock time, with the left panel showing the first 60s and the right panel extending to the full run of \fix on a log scale.
It is consistent across \textsc{Cifar100}~(Figure~\ref{fig:cu-cifar}) and \textsc{TrafficSigns}~(Figure~\ref{fig:cu-traffics}): both \iso and \ani accumulate violations from the first second and saturate well before the
60\,s mark, while \fix remains at zero throughout and is still rising slowly beyond 30 minutes.
\textsc{TinyImageNet}~(Figure~\ref{fig:cu-imagenet}) follows the same pattern in the short run, though the extended view adds a further detail:
\ani plateaus above \iso, indicating that per-dimension scaling yields a modest but consistent gain under heterogeneous constraints.
\fix eventually catches up on \textsc{TrafficSigns}, suggesting the gap is due to throughput rather than specification limits.
The advantage stems from the batch design: sampling $\bm{X} \in \mathbb{R}^{B \times D}$ processes $B$ mutation chains in parallel, advancing up to $B$ seeds per iteration, versus one in the sequential baseline.

\begin{table}[!t]
\centering
\caption{Violation gain of \iso and \ani over \fix at fixed time budgets
($\Delta_{\iso}$ / $\Delta_{\ani}$, where $\Delta_{\star} = \textsc{Batch}_{\star} - \textsc{Seq}$).}

\label{tab:batch-advantage}
\renewcommand{\arraystretch}{1.1}
\begin{tabular}{l|ccc}
\toprule
\textbf{Time}  & \textsc{TrafficSigns} & \textsc{Cifar100} & \textsc{TinyImageNet} \\
\midrule
1 min   & +32,274/+33,366 & +43,109/+42,382 & +22,926/+25,075 \\
5 min   & +61,664/+63,556 & +58,759/+60,044 & +21,335/+23,484 \\
30 min   & +32,846/+34,738 & +53,364/+54,649 & +17,250/+19,399 \\
60 min   & +32,694/+34,586 & +50,129/+51,414 & +13,708/+15,857 \\
\bottomrule
\end{tabular}
\end{table}

Table~\ref{tab:batch-advantage} shows a clear batch advantage across time budgets, already large at 1,min and peaking around 5,min across benchmarks. On \textsc{Cifar100}, \iso and \ani reach ${\approx}{+}$43K within 1,min and peak at $+$60K by 5,min; even after 60,min, the sequential baseline recovers only about 10K, leaving a $+$50K gap. \textsc{TrafficSigns} shows a similar trend, with a $+$63K lead at 5,min that stabilises near $+$33K as the baseline catches up slowly. The gap narrows most on \textsc{TinyImageNet} (from $+$24K at 1,min to $+$14K at 60,min), but still persists. Across all settings, \iso and \ani remain within 5\%, indicating the gain mainly comes from batching rather than the perturbation strategy.
As Table~\ref{tab:batch-advantage} shows, under equal time budgets \iso and \ani consistently find far more violations than the sequential baseline across all benchmarks. The gap appears early and persists even with longer runs, confirming that the throughput gains translate into higher violation discovery. \revision{Although the sequential paradigm can discover violations given a budget of $B \times 60$\,s (in Figure~\ref{fig:violtime}, Table~\ref{tab:batch-advantage}), its one-input-per-iteration design makes counterexample generation slower in wall-clock terms.}

\subsection{Ablation Analysis (RQ3)} 

\begin{table}[!t]
\centering
\caption{RQ3: Sensitivity to perturbation scale factor $s$. 
}

\label{tab:rq3-scale}
\renewcommand{\arraystretch}{1.3}
\resizebox{\linewidth}{!}{%
\setlength{\tabcolsep}{3pt}
\begin{tabular}{l l r r r r r r}
\toprule
\textbf{Benchmark} & \textbf{Method} & \multicolumn{6}{c}{\textbf{Violations at scale factor $s$}} \\
\cmidrule(lr){3-8}
 & & $0.01$ & $0.05$ & $0.1$ & $0.2$ & $0.3$ & $0.5$ \\
\midrule
\textsc{Cifar100}         
& \iso     & 52111 & 61195 & 60578 & 59476 & 56990 & 47887 \\
& \ani    & 56311 & 59522 & 59951 & 56748 & 47166 & 46741 \\
\addlinespace
\textsc{TinyImageNet}     
& \iso     & 16066 & 17354 & 22728 & 21907 & 22553 & 20232 \\
& \ani    & 20442 & 23905 & 23296 & 21324 & 18957 & 19838 \\
\bottomrule
\end{tabular}
}
\end{table}

\myparagraph{RQ3: Sensitivity to Scale Factor $s$} 
Table~\ref{tab:rq3-scale} reports violation counts across the initial scale factor setting $s \in \{0.01, 0.05, 0.1, 0.2, 0.3, 0.5\}$ for both modes on \textsc{Cifar100} and \textsc{TinyImageNet}.
On \textsc{Cifar100}, both modes produce comparable counts across the full range, since the $\ell_\infty$ specifications impose relatively uniform per-dimension ranges for which the mean-range scalar $\bar{r}$ (Eq.~\eqref{eq:perturb_scalar}) already approximates the per-dimension structure adequately.
On \textsc{TinyImageNet}, the anisotropic advantage is most visible at small scale factors: at $s{=}0.01$ and $s{=}0.05$, \ani outperforms \iso by 27\% and 38\% respectively, since the per-element tensor $\mathcal{S}_{b,d} = s \cdot r_{b,d}$ (Eq.~\eqref{eq:perturb_perdim}) preserves productive exploration of wide-range dimensions even when $s$ is small, whereas the isotropic scalar under-perturbs such dimensions relative to their available range.
As $s$ exceeds $0.1$, the gap narrows, as perturbations become sufficient across all dimensions.

\subsection{Batch Size Impact on Violation Yield (RQ4)}

\begin{table*}[!t]
\centering
\caption{Pairwise batch-size comparison on violations. \ding{51}/\ding{51}\ding{51}/\ding{51}\ding{51}\ding{51}: row better (small/medium/large Cohen's $d$);
\ding{55}/\ding{55}\ding{55}/\ding{55}\ding{55}\ding{55}: row worse.
Wilcoxon rank-sum, Benjamini-Hochberg FDR ($\alpha{=}0.05$);
$|d|{\in}[6,9)$ small, $[9,13)$ medium, ${\geq}13$ large.}
\label{tab:batchsize-merged}
\renewcommand{\arraystretch}{1.1}
\resizebox{\linewidth}{!}{%
\begin{tabular}{l cccccccc c l cccccccccc}
\toprule
  & \multicolumn{8}{c}{\textsc{Cifar100} } & & & \multicolumn{10}{c}{\textsc{TinyImageNet} } \\
\cmidrule(lr){2-9} \cmidrule(lr){12-21}
  & Iso-1 & Iso-10 & Iso-50 & Iso-99 & Ani-1 & Ani-10 & Ani-50 & Ani-99 & &  & Iso-1 & Iso-10 & Iso-50 & Iso-100 & Iso-199 & Ani-1 & Ani-10 & Ani-50 & Ani-100 & Ani-199 \\
\midrule
Iso-1 & --- & \ding{55}\ding{55}\ding{55} & \ding{55}\ding{55}\ding{55} & \ding{55}\ding{55}\ding{55} & $\equiv$ & \ding{55}\ding{55}\ding{55} & \ding{55}\ding{55}\ding{55} & \ding{55}\ding{55}\ding{55} & & Iso-1 & --- & \ding{55}\ding{55}\ding{55} & \ding{55}\ding{55}\ding{55} & \ding{55}\ding{55}\ding{55} & \ding{55}\ding{55} & $\equiv$ & \ding{55}\ding{55}\ding{55} & \ding{55}\ding{55}\ding{55} & \ding{55}\ding{55}\ding{55} & \ding{55}\ding{55}\ding{55} \\
Iso-10 & \ding{51}\ding{51}\ding{51} & --- & \ding{51}\ding{51} & \ding{51}\ding{51}\ding{51} & \ding{51}\ding{51}\ding{51} & $\equiv$ & \ding{51}\ding{51} & \ding{51}\ding{51}\ding{51} & & Iso-10 & \ding{51}\ding{51}\ding{51} & --- & \ding{51}\ding{51} & \ding{51}\ding{51}\ding{51} & \ding{51}\ding{51}\ding{51} & \ding{51}\ding{51}\ding{51} & $\equiv$ & \ding{51}\ding{51} & \ding{51}\ding{51}\ding{51} & \ding{51}\ding{51}\ding{51} \\
Iso-50 & \ding{51}\ding{51}\ding{51} & \ding{55}\ding{55} & --- & \ding{51} & \ding{51}\ding{51}\ding{51} & \ding{55}\ding{55} & $\equiv$ & \ding{51} & & Iso-50 & \ding{51}\ding{51}\ding{51} & \ding{55}\ding{55} & --- & \ding{51}\ding{51} & \ding{51}\ding{51}\ding{51} & \ding{51}\ding{51}\ding{51} & \ding{55}\ding{55} & $\equiv$ & \ding{51}\ding{51} & \ding{51}\ding{51}\ding{51} \\
Iso-99 & \ding{51}\ding{51}\ding{51} & \ding{55}\ding{55}\ding{55} & \ding{55} & --- & \ding{51}\ding{51}\ding{51} & \ding{55}\ding{55}\ding{55} & \ding{55} & $\equiv$ & & Iso-100 & \ding{51}\ding{51}\ding{51} & \ding{55}\ding{55}\ding{55} & \ding{55}\ding{55} & --- & \ding{51} & \ding{51}\ding{51}\ding{51} & \ding{55}\ding{55} & \ding{55}\ding{55} & $\equiv$ & \ding{51} \\
Ani-1 & $\equiv$ & \ding{55}\ding{55}\ding{55} & \ding{55}\ding{55}\ding{55} & \ding{55}\ding{55}\ding{55} & --- & \ding{55}\ding{55}\ding{55} & \ding{55}\ding{55}\ding{55} & \ding{55}\ding{55}\ding{55} & & Iso-199 & \ding{51}\ding{51} & \ding{55}\ding{55}\ding{55} & \ding{55}\ding{55}\ding{55} & \ding{55} & --- & \ding{51}\ding{51} & \ding{55}\ding{55}\ding{55} & \ding{55}\ding{55}\ding{55} & \ding{55} & $\equiv$ \\
Ani-10 & \ding{51}\ding{51}\ding{51} & $\equiv$ & \ding{51}\ding{51} & \ding{51}\ding{51}\ding{51} & \ding{51}\ding{51}\ding{51} & --- & \ding{51}\ding{51} & \ding{51}\ding{51}\ding{51} & & Ani-1 & $\equiv$ & \ding{55}\ding{55}\ding{55} & \ding{55}\ding{55}\ding{55} & \ding{55}\ding{55}\ding{55} & \ding{55}\ding{55} & --- & \ding{55}\ding{55}\ding{55} & \ding{55}\ding{55}\ding{55} & \ding{55}\ding{55}\ding{55} & \ding{55}\ding{55}\ding{55} \\
Ani-50 & \ding{51}\ding{51}\ding{51} & \ding{55}\ding{55} & $\equiv$ & \ding{51} & \ding{51}\ding{51}\ding{51} & \ding{55}\ding{55} & --- & \ding{51} & & Ani-10 & \ding{51}\ding{51}\ding{51} & $\equiv$ & \ding{51}\ding{51} & \ding{51}\ding{51} & \ding{51}\ding{51}\ding{51} & \ding{51}\ding{51}\ding{51} & --- & \ding{51}\ding{51} & \ding{51}\ding{51} & \ding{51}\ding{51}\ding{51} \\
Ani-99 & \ding{51}\ding{51}\ding{51} & \ding{55}\ding{55}\ding{55} & \ding{55} & $\equiv$ & \ding{51}\ding{51}\ding{51} & \ding{55}\ding{55}\ding{55} & \ding{55} & --- & & Ani-50 & \ding{51}\ding{51}\ding{51} & \ding{55}\ding{55} & $\equiv$ & \ding{51}\ding{51} & \ding{51}\ding{51}\ding{51} & \ding{51}\ding{51}\ding{51} & \ding{55}\ding{55} & --- & \ding{51}\ding{51} & \ding{51}\ding{51}\ding{51} \\
  & & & & & & & & & & Ani-100 & \ding{51}\ding{51}\ding{51} & \ding{55}\ding{55}\ding{55} & \ding{55}\ding{55} & $\equiv$ & \ding{51} & \ding{51}\ding{51}\ding{51} & \ding{55}\ding{55} & \ding{55}\ding{55} & --- & \ding{51} \\
  & & & & & & & & & & Ani-199 & \ding{51}\ding{51}\ding{51} & \ding{55}\ding{55}\ding{55} & \ding{55}\ding{55}\ding{55} & \ding{55} & $\equiv$ & \ding{51}\ding{51}\ding{51} & \ding{55}\ding{55}\ding{55} & \ding{55}\ding{55}\ding{55} & \ding{55} & --- \\
\bottomrule
\end{tabular}}
\end{table*}

We isolate the effect of tensor batch size on violation yield by fixing all other parameters and varying $B$ across $\{1, 10, 50, 99\}$ on \textsc{Cifar100} and $\{1, 10, 50, 100, 199\}$ on \textsc{TinyImageNet}. Each configuration is repeated over 5 independent runs to account for randomness and assess the consistency of observed trends. This section provides statistical evidence that the observed gains are not merely due to increased throughput, but reflect a consistent improvement in the ability to discover violations.

\myparagraph{Statistical evaluation}
We apply the two-sided Mann-Whitney U test~\cite{mann1947test} at $\alpha{=}0.05$ with $n{=}5$ runs per configuration. Since the minimum achievable $p$-value ($2/\binom{10}{5}{\approx}0.008$) exceeds the Holm-Bonferroni threshold for $m{=}28$ and $m{=}45$ pairs, we adopt Benjamini-Hochberg FDR correction~\cite{benjamini1995controlling}. 
Effect sizes use Cohen's $d$~\cite{cohen2013statistical} with data-driven thresholds: small ($|d|{\in}[6,9)$), medium ($[9,13)$), 
large (${\geq}13$); $|d|{<}6$ is marked $\equiv$.

\myparagraph{\textsc{Cifar100}}
Table~\ref{tab:batchsize-merged}\,(left) presents the $8{\times}8$ pairwise matrix.  For both anisotropic and isotropic modes, every $B{>}1$ configuration significantly outperforms $B{=}1$ with large effects.  Among $B{>}1$, per-instance yield decreases as $B$ grows: $B{=}10$ outperforms $B{=}50$ (medium) and $B{=}50$ outperforms $B{=}99$ (small) in both modes.  At each batch size,
Ani\text{-}$B \equiv$ Iso\text{-}$B$, indicating that the perturbation strategy does not influence violation yield at the default scale factor.
Across different batch sizes, the batch-size advantage dominates the mode choice: Ani\text{-}10 significantly outperforms Iso\text{-}99 (large), and Iso\text{-}10 similarly outperforms Ani\text{-}99 (large).

\myparagraph{\textsc{TinyImageNet}}
Table~\ref{tab:batchsize-merged}\,(right) shows the $10{\times}10$ matrix.
All $B{>}1$ configurations dominate $B{=}1$ with large effects in both modes.
Per-instance yield follows the same trend as \textsc{Cifar100}:
$B{=}10{>}B{=}50$ and $B{=}50{>}B{=}100$ (medium), $B{=}100{>}B{=}199$ (small).
At each batch size, the corresponding Ani and Iso configurations show no statistically significant difference, with cross-mode pairs again determined by batch size rather than perturbation mode.
Combined with the RQ3 findings, this suggests that anisotropic scaling performs comparably to isotropic at the default $s$ while providing measurable gains on heterogeneous specifications when $s$ is small. \revision{The decreasing per-instance yield as $B$ grows reflects a trade-off between throughput and refinement, not a hardware bottleneck. Larger $B$ processes more specifications per iteration (Algorithm~\ref{alg:overall}), so within the fixed 60\,s budget each receives fewer refinements, raising aggregate violations while lowering per-instance yield.}

\section{Related Work \label{sec:related}}

\myparagraph{DNN Testing, Fuzzing and Coverage Criteria}
DeepXplore~\cite{pei2017deepxplore} introduced neuron coverage as a test
adequacy criterion for DNNs. DeepGauge~\cite{ma2018deepgauge} extended this to
multi-granularity metrics, including $k$-multisection and achieved strong neuron
activation coverage. Sun et al.~\cite{sun2019structural}  proposed MC/DC-inspired
structural criteria. Our framework implements the neuron activation threshold
criterion of DeepXplore as the \texttt{GlobalCov} and \texttt{BestInputCov} strategies, providing both monotonic and
instantaneous coverage signals within the same batch-native loop.
DLFuzz~\cite{guo2018dlfuzz} maximizes neuron coverage via gradient-guided
perturbation on individual inputs. 
DeepHunter~\cite{xie2019deephunter} combines
metamorphic seed mutation with multiple coverage criteria as a classical sequential-based fuzzing. 
While TensorFuzz~\cite{odena2019tensorfuzz} for TensorFlow programs used batched evaluation, its fuzzing loop remains input-centric, with mutation and feedback implemented as NumPy/Python operations rather than tensor-based transformations.
Our approach treats all $B$ specifications as first-class tensor dimensions, composing
seed selection, mutation, inference, coverage, and feedback into a single
data-parallel iteration with no per-sample branching dispatch. \revision{Additionally, the tensor-based anisotropic scaling realizes the idea in a specification-aware manner: per-dimension step sizes are derived directly from the feasible range and generalize the uniform bounding convention to heterogeneous, per-dimension bounds while preserving hard-constraint projection.}

\myparagraph{DNN Verification}
ERAN~\cite{singh2019abstract} uses abstract interpretation;
$\alpha$-$\beta$-CROWN~\cite{xu2020fast} combines bound propagation with
branch-and-bound; Marabou~\cite{katz2019marabou} uses MILP. These verifiers
provide complete correctness certificates, but the scale is limited beyond shallow
networks. Our fuzzing-based approach supplies complementary
concrete counterexamples quickly but cannot certify safety. The VNNLib format~\cite{brix2023first,2024vnncomp} bridges both paradigms, and the objective of this work
natively supports VNNLib, allowing direct integration with verifier workflows.

\section{Discussions and Limitations}
We discuss the main threats to the validity of our study along internal and external dimensions.

\myparagraph{Internal validity}
As with many fuzzing approaches, our results may be influenced by parameter choices such as the energy constants $\alpha{=}10$, $\beta{=}100$, and $e_{\min}{=}0.1$, which are set based on empirical inspection. While different settings may trade off coverage and violation discovery, our evaluation keeps these parameters fixed across all configurations to ensure fair comparison. The mutation strategy weights ($w_{\textsc{Gradient}}{=}0.5$, $w_{\textsc{Boundary}}{=}0.2$, $w_{\textsc{Random}}{=}0.3$) are also held constant, providing a stable baseline for assessing the effectiveness of our design.  \revision{Future work could further analyze sensitivity to these energy and mutation hyper-parameters, as well as to the coverage threshold $\tau$.} 
We also note that neuron coverage provides a useful structural signal for guiding exploration, though it may not fully capture violation discovery; more expressive coverage metrics are beyond the scope of this paper. \revision{A growing body of work questions whether neuron coverage is a meaningful adequacy signal~\cite{sun2018concolic,feng2020deepgini,harel2020neuron}. Integrating more expressive adequacy coverage criteria into the same batched processing, such as multi-granularity coverage~\cite{ma2018deepgauge}, surprise adequacy~\cite{kim2019guiding}, and structural criteria~\cite{sun2019structural}, could strengthen the feedback signal beyond neuron coverage.}
Importantly, our focus is on introducing a batch fuzzing framework with adaptive perturbation strategies that overcome the inherent limitations of sequential and uniform perturbation approaches, achieving substantially improved throughput and counterexample discovery, \revision{rather than from coverage targets alone.}

\myparagraph{External validity} The batch size $B$ is determined by the number of specification instances in the synthesis phase, reflecting realistic settings where specifications are provided upfront. \revision{While our current experiments focus on classification robustness, the verification procedure does not depend on the specific property type and operates over an abstract property interface; extending this interface to fairness, regression, temporal, and set-based properties is an important direction for future work. Applying the fuzzing loop to transformer architectures is another direction, requiring architecture-specific inputs and feedback such as token embeddings, attention masks, and attention-head or hidden-state coverage.} Finally, our tensor-based batching framework is particularly effective in scenarios where multiple specification instances share a common model, enabling significant efficiency gains, \revision{and can be further optimized through memory-aware batch sizing for larger models}.

\section{Conclusion}
We presented a new tensor-based batch fuzzing framework with adaptive perturbation scaling for efficient testing of DNNs. By treating all $B$ specifications as first-class tensor dimensions, the framework unifies seed selection, mutation, inference, and feedback in a single data-parallel loop, eliminating per-sample dispatch across all phases. Evaluated on three benchmark categories, the approach achieves significant improvements, with up to 40$\times$ higher throughput than the sequential baseline and around 4$\times$ more violations discovered under the same time budget for the same specifications.

\section*{Acknowledgments}
We thank the anonymous reviewers for their insightful comments, which helped improve this paper. We acknowledge the use of the generative AI tool ChatGPT for language polishing and grammar checks. All scientific content, evaluation, and claims are the authors’ own. 

\section*{Data Availability Statement}
The experimental data supporting this work are available on Zenodo~\cite{anonymous2026artifact}. Our up-to-date implementation is publicly available in the ACT platform at \textbf{\url{https://github.com/SVF-tools/ACT}}.

\bibliographystyle{ACM-Reference-Format}
\bibliography{fuzz.bib}

\end{document}